%% file: samplepaper.tex
\documentclass[runningheads]{llncs}
\usepackage{graphicx}

\usepackage{cite}
\usepackage{amsmath,amssymb,amsfonts}
\usepackage{algorithmic}
\usepackage{graphicx}
\usepackage{textcomp}
\usepackage{xcolor}
\usepackage{graphicx}
\usepackage[linesnumbered,ruled,algonl,vlined,noend]{algorithm2e} 
\usepackage{graphicx}
\usepackage{tabularx}
\usepackage{booktabs}
\usepackage{amssymb, amsmath}
\usepackage{float}
\usepackage{subfig}
\usepackage{wrapfig}
\newtheorem{defn}{Definition}
\newtheorem{thm}{Theorem}
\newtheorem{lem}{Lemma}
\begin{document}
\title{Proactive Rumor Control: When Impression Counts (Full Version)}
%
%\titlerunning{Abbreviated paper title}
% If the paper title is too long for the running head, you can set
% an abbreviated paper title here
%
\author{Pengfei Xu\and
Zhiyong Peng \and
Liwei Wang}
%
%\authorrunning{Pengfei Xu et al.}
% First names are abbreviated in the running head.
% If there are more than two authors, 'et al.' is used.
%
\institute{ School of Computer Science, Wuhan University, Hubei, China\\
\email{\{xupengfei,peng,liwei.wang\}@whu.edu.cn}}
\maketitle              % typeset the header of the contribution
\input{./tex/macros}

\begin{abstract}
The spread of rumors in online networks threatens public safety and results in economic losses. To overcome this problem, a lot of work studies the problem of rumor control which aims at limiting the spread of rumors. However, all previous work ignores the relationship between the influence block effect and counts of impressions on the user.
In this paper, we study the problem of minimizing the spread of rumors when impression counts. Given a graph $G(V,E)$, a rumor set $R \in V$ and a budget $k$, it aims to find a protector set $P \in V \backslash R$ to minimize the spread of the rumor set $R$ under the budget $k$. Due to the impression counts, two following challenges of our problem need to be overcome: (1) our problem is NP-hard; (2) the influence block is non-submodular, which means a straightforward greedy approach is not applicable. Hence, we devise a branch-and-bound framework for this problem with a ($1-1/e-\epsilon$) approximation ratio. To further improve the efficiency, we speed up our framework with a progressive upper bound estimation method, which achieves a ($1-1/e-\epsilon - \rho$) approximation ratio. We conduct experiments on real-world datasets to verify the efficiency, effectiveness, and scalability of our methods.

\end{abstract}

\input{./tex/intro}

\input{./tex/related}

\input{tex/problem}

\input{./tex/Ourframework}

\input{./tex/pro}

\input{./tex/exp}

\input{./tex/con}

\bibliographystyle{splncs04}
\bibliography{cas-refs}
\end{document}

%% file: tex/macros.tex
\newcommand{\prob}{RCIC}

\newcommand{\smallVspace}{\vspace{-0.1in}}
\newcommand{\bigVspace}{\vspace{-0.2in}}

\newcommand{\probinf}{CIM}

\newcommand{\wk}{w} 
\newcommand{\pos}{vt} 
\newcommand{\object}{\mathcal{G}} 
\newcommand{\qmatrix}{\mathbf{Q}}

\newcommand{\forwardlist}{\mathcal{F}}
\newcommand{\blockdegree}{\mathcal{D}}
\newcommand{\mappingIndex}{\mathcal{B}}
\newcommand{\listG}{num}

\newcommand{\Wu}{\ensuremath{{w_u}}}                   %%a random walk start from u
\newcommand{\puv}{\ensuremath{p_{uv}}}                 %% the probability of u pick it's neighbor v

\newcommand{\pr}{\ensuremath{{P}}} 
\newcommand{\ru}{\ensuremath{{R}}} 
\newcommand{\IS}{\ensuremath{{Is}}}
\newcommand{\ic}{\ensuremath{{C}}}
\newcommand{\I}{\ensuremath{{I}}}
\newcommand{\OPT}{\ensuremath{O}}

\newcommand{\getInfluenceSetR}{getInfluenceSet}
\newcommand{\getBlockSetv}{getBlockSet}
\newcommand{\getVisitedNodes}{getVisitedNodes}
\newcommand{\updateBlockSetv}{updateBlockSet}

\newcommand{\degreeTop}{TopK}
\newcommand{\samgreedy}{Greedy}
\newcommand{\bab}{BranchAndBound}
\newcommand{\sbound}{SamComputeBound}
\newcommand{\psbound}{ProSamComputeBound}
\newcommand{\mbound}{MatComputeBound}
\newcommand{\samgreedyplus}{GreedySel*}
\newcommand{\indexgreedyplus}{IndexSel*}
\newcommand{\rankgreedy}{RanSel} 
\newcommand{\fastrank}{FastRank}

%% file: tex/intro.tex
\section{Introduction}
World Wide Web and social networks have become the most commonly utilized vehicles for information propagation and changed people's lifestyles greatly due to the increasing popularity of online networks. However, the ease of information propagation is a double-edged sword. But rumors and misinformation could be quickly spread on social networks, which results in undesirable social effects and even leads to economic losses~\cite{DBLP:conf/www/BudakAA11,DBLP:conf/cikm/TripathyBM10}.%\cite{DBLP:conf/websci/NguyenYTE12,DBLP:conf/www/BudakAA11,DBLP:conf/csonet/ZhangZLT15,DBLP:conf/cikm/TripathyBM10}. %For example, the rumor of explosion in the White House in 2013 caused \$130 billion loss on the stock market\footnote{https://www.dailymail.co.uk/sciencetech/article-3090221} and the fake tweet about the earthquake in Ghazni province in August 2012 made thousands of people leave their home in a panic and be afraid of returning home~\cite{DBLP:conf/icdcs/FanLWTMB13}. 
Therefore, minimizing the spread of rumors in online networks is a crucial problem.

To solve this problem, a lot of work studies the problem of rumor control which aims to minimize the spread of rumors on social network~\cite{DBLP:conf/wine/BharathiKS07,DBLP:conf/wine/BorodinFO10,DBLP:conf/ACMicec/CarnesNWZ07,albert2000error,newman2002email,DBLP:conf/kdd/HabibaYBS08}.
%in this work, we consider a strategy that initiates protectors to fight against rumors on social networks. 
%Unfortunately, most previous work 
However, they only assume that users are passive receivers of rumors even if the users can browse the rumors on their own. Therefore, in this study, we assume that users will actively encounter/contact the rumors via their browsing behaviors, i.e., keyword search, social browsing, etc, which can be modeled by random walk model~\cite{DBLP:journals/www/ZhangBNZMGP19,DBLP:conf/nctcs/MoT0P19}. Unfortunately, existing work~\cite{DBLP:journals/www/ZhangBNZMGP19,DBLP:conf/nctcs/MoT0P19} does not consider the relationship between the influence block and counts of impressions on one user because the model assumes one-time impression is enough. But in the real world, studies in consumer behavior report that users are unlikely to take meaningful action when they receive a message only one time~\cite{feder1985adoption,lancaster1990econometric}. Meanwhile, there is evidence showing that the effect of message repetition should be measured as an S-shaped function (logistic function)~\cite{palda1965measurement,taylor2009once}.

To this end, we study the problem of minimizing the spread of rumor
when impression counts and call it Rumor Control when Impression Counts
(\prob). Suppose that an online network is represented by a graph $G(V,E)$. 
Given a rumor set $R \in V$ and a budget $k$, \prob~aims to find a protector set $P \in V \backslash R$ to minimize the spread of the rumor set $R$ as much as possible under the budget $k$. To the best of our knowledge, this is the first problem for rumor control when impression counts are considered. As a result, the following challenges are important to be addressed.

The first challenge is the NP-hardness of \prob~as we analyze in Theorem~\ref{NP-hardness}. Then, we resort to developing approximate algorithms to solve it efficiently. The second challenge is posed by the property of the logistic function. The influence block model based on the logistic function is non-submodular, which means any straightforward greedy-based
the approach is not applicable to address the \prob~problem as shown in Example~\ref{non-sub}. To overcome this challenge, we proposed a sampling-based greedy method to estimate the upper bounds of the logistic function value. Based on this upper bound estimating method, we devise
a branch-and-bound framework for \prob, with a ($1-1/e-\epsilon$) approximation ratio. Furthermore, we speed up our framework with a progressive upper-bound estimation method.
%However, this framework still suffers from a high computational cost due to heavily invoking the upper bound estimation method. Therefore, to further improve the efficiency of our framework, we devise a progressive sampling-based greedy method for the upper bound estimation, which can provide a trade-off between efficiency and effectiveness.
In summary, we make the following contributions.
\begin{itemize}
	\item We propose and study the \prob~problem, and analyze the monotonicity and non-submodularity of the objective function of \prob. We show that \prob~is NP-hard.
	\item To solve the \prob~problem, we present a Monte Carlo based greedy algorithm (\samgreedy) as the baseline solution. Moreover, we devise an upper-bound estimation method by adaptively solving submodular optimization problems. Based on the upper bound function, we propose a branch-and-bound framework for \prob, with a ($1-1/e-\epsilon$) approximation ratio.
	\item To further improve the efficiency, we speed up our framework with a progressive sampling-based greedy method for upper bound estimation, which achieves a ($1-1/e-\epsilon - \rho$) approximation ratio and a significant reduction in running time.
	\item We conduct extensive experiments on three real-world datasets. The results validate the effectiveness, efficiency, and scalability of our solutions.
\end{itemize}

%% file: tex/related.tex
\section{Related work}
In the following, we discuss the most relevant literature to our problem.

Two proactive rumor control problems in online networks are close to our work~\cite{DBLP:journals/www/ZhangBNZMGP19,DBLP:conf/nctcs/MoT0P19}, which also study proactive rumor control problem to minimize the spread of the rumor set under the budget. The core difference lies in the influence block model. In particular, the existing work assumes that one anti-rumor node before the rumor node can block the total influence of the rumor set to the user in one browsing process. %Under such an influence block model, when multiple billboards are close to a trajectory, the marginal influence is reduced to capture the property of diminishing returns. Therefore, TIP focuses on identifying and reducing the overlap of the influence among different billboards to the same trajectories while keeping the budget constraint into consideration. That is, TIP can maximize the number of distinct users by impressing as many people as possible for one time. 
It does not consider the relationship between the influence block effect and counts
of impressions on one user because the model assumes one time
impression is enough. On the contrary, \prob~is built upon a logistic influence block model, which has been widely adopted in consumer behavior studies. To minimize the spread of the rumor set, we need to control the overlap to some extent by impressing the same users several times. %Unfortunately, the logistic influence model is non-submodular. Adapting the greedy approach to ICOA, which effectively solves TIP, could lead to arbitrarily bad solutions due to the non-submodular of the influence function.
%, which only studies one specific case of our problem, assuming that the cost of all nodes in $G$ is uniform. Nevertheless, in fact, we cannot ignore the cost information of the nodes. For example, it is impossible that the price of broadcasting information on Baidu's homepage is the same as personal homepage's. Therefore, we study the rumor control within budget constraint problem.

Two other problems close to our problem are influence block and competitive influence maximization. 
Influence block aims to limit the influence of rumors by blocking some nodes or links in a network~\cite{albert2000error,newman2002email,DBLP:conf/kdd/HabibaYBS08}. Their strategies of the seed selection are mainly based on their connectivity, such as degree~\cite{albert2000error,newman2002email}, pagerank~\cite{DBLP:conf/kdd/HabibaYBS08}, and
betweenness~\cite{DBLP:conf/kdd/HabibaYBS08}.
Different from the first problem, competitive influence maximization tries to identify a set of target seed nodes (or protectors) who will spread an `anti-rumor’ to limit the scale of rumor propagation~\cite{DBLP:conf/wine/BharathiKS07,DBLP:conf/wine/BorodinFO10,DBLP:conf/ACMicec/CarnesNWZ07}. Carnes et al.~\cite{DBLP:conf/ACMicec/CarnesNWZ07}, and Bharathi et
al.~\cite{DBLP:conf/wine/BharathiKS07} study competitive influence diffusion under the extension of the Independent Cascade model and show
that the problem of maximizing the influence of one campaign is NP-hard and submodular,
while Borodin et al.~\cite{DBLP:conf/wine/BorodinFO10} studies the similar problem under the Linear Threshold model. 
Our problem is essentially different from the above work for the following reason. Both influence block and competitive influence maximization assume that the information (or rumors) propagations are driven by the effect of word-of-mouth, and they use the Independent Cascade model and Linear Threshold model to simulate the spread of rumors. However, our problem assumes that rumors spread via browsing behaviors and uses a random walk model to describe the influence spread of rumors.

%% file: tex/problem.tex
\section{Problem formulation}\label{formulation}
In this section, we first formally define the influence model and influence block. In the following, we give the formulation definition of \prob.  In the end, we show the non-submodularity of the objective function of \prob~and prove that \prob~is NP-hard. %Important notations used in our paper are presented in Table~\ref{table:notation}.
% \vspace{-0.2cm}
% \begin{table*}[t]
% 	\caption{{Notations for problem formulation and solutions}}
% 	\centering
% 	\begin{tabular}{lp{9.0cm}}
% 		\toprule
% 		{\bfseries Symbol} & {\bf Description}\\
% 		\midrule
% 		$G(V,E)$ & An online graph \\
% 		$\puv$ & the transition probability from neighbor node $u$ to node $v$\\
% 		$\wk_u$ & An instance of a random walk starting from $u$\\
% 		$T$     & A given threshold to bound the length of $\wk_{u}$\\
% 		$k$     & A given budget\\
% 		$R$, $P$    & $R \subset V$ is a rumor set, $P \subset V$ is a protector set, and $R \cap P=\phi$  \\
% 		$\ic_{\Wu}(\pr|\ru)$ &the total impressions of $P$ to block the influence of $R$ to $u$ in $\Wu$\\
% 		$I_{\Wu}(\pr|\ru)$ &The value of probability that $P$ blocks the influence of $R$ to $u$ for $\wk_u$\\
% 		$I_{u}(\pr|\ru)$  & $ I_{u}(\pr|\ru) = E[I_{\Wu}(\pr|\ru)]$ for any $\wk_u$\\ 
% 		$\object({\pr}|{\ru})$ & The object function of our problem (the block degree of $P$)\\
% 		\bottomrule
% 	\end{tabular}
% 	\label{table:notation}
% \end{table*}
\vspace{-0.3cm}
\subsection{Influence model}\vspace{-0.2cm}
Let $G = (V ,E)$ be an online network with $n = |V|$ nodes and $m = |E|$ edges. The random walk process can be used to model the user’s browsing process on $G$ as follows~\cite{spitzer2013principles,DBLP:journals/pvldb/MoBZP20a,DBLP:journals/www/ZhangBNZMGP19,DBLP:conf/nctcs/MoT0P19}. Given a node $u \in V$, a browsing process starting from $u$ can be represented by a random walk $\Wu$. In particular, $w_u$ picks a neighbor $v$ of $u$ by the probability of $\puv=1/|$neighbors of $v|$ and moves to this neighbor and then follows this way recursively. We say that $u$ hits $v$ at step $t$, if $\wk_u$ first visits $v$ after $t$ walk steps.

%This process can be viewed as a Markov chain, and each element $\puv$ in the transition matrix $\pr$ is given by:
% \begin{equation}
% {p_{uv}} = \left\{ {\begin{array}{*{20}{l}}
% 	{1/d_u}	&	{\textrm{$u$ is a neighbour of $v$}}\\
% 	0			&	\textrm{otherwise}
% 	\end{array}} \right.
% \end{equation}
% where $p_{uv}$ is the transition probability from $u$ to $v$, and $d_u$ is the out-degree of~$u$.

%Specifically, given an arbitrary random walk $\wk_u=(n_0, n_1,...)$, where $(...)$ represents a sequence of nodes, and $u=n_0$ starting from $n_0$ and a node $v \in V$. We say that
% $u$ hits $v$ at step $t$, if $\wk_u$ first visits $v$ after $t$ walk steps. 
% We define that $\wk_{uv}$ is a hitting path of $u$ to $v$, if $\wk_{uv}=(n_0, n_1,...,n_k)$, $u=n_0$, and $v=n_k$, for $v \neq n_0, n_1...,n_{k-1}$. Let $\mathcal{W}_{uv}$ denote the set of all possible $\wk_{uv}$, the possibility of $\wk_{uv}$ being generated is computed by $\mathcal{P}(\wk_{uv})=\prod\nolimits_{i = 0}^{k - 1} 1/{{d_i}} $.

Similarly, we say that $u$ hits (or is influenced by) set $S$ at the time step $t$ if $\Wu$ first visits set $S$ by a $t$-hop jump. It is worth noting that $t$ should not be very large in the real world, as most social media users only browse a small number of pages each day. Therefore, we can use a threshold $T$ to bound the hitting time $t$ for any nodes and sets.

\subsection{Influence block}
Based on the influence model, we introduce the concept of influence block when impression counts as follows. 

Before that, we first introduce the conception of impression. For nodes $n_1$ and $n_2$ in a random walk $\Wu$, we define that $n_1$ have an impression of blocking the influence of $n_2$ to $u$ if $\Wu$ visits $n_1$ before $n_2$.
%w.r.t. $\Wu$ first visits $n_1$ and $n_2$ at the time step $t_1$ and $t_2$ respectively, and $t_1 < t_2$. 
Therefore, we use the Bernoulli random variable $\ic_{\Wu}(n_1|n_2)$ denoting the states whether $n_1$ have a impression of block the influence of $n_2$ to $u$, where $\ic_{\Wu}(n_1|n_2) = 1$ denotes that $n_1$ have a impression of block the influence of $n_2$ to $u$, otherwise $\ic_{\Wu}(n_1|n_2) = 0$. Then the total impressions of $P$ ($P \subset V$ is a protector set) to block the influence of $R$ ($R \subset V$ is a rumor set) to $u$ in $\Wu$ can be computed by $\ic_{\Wu}(P|R) = \sum\nolimits_{v \in \pr} {\ic_{\Wu}(v|n_r)}$. Here $n_r$ is the first node in $\Wu$, which is contained in the set $R$. 

Our influence block when impression counts are based on the logistic function. We use the following equation to compute the influence block of a protector set $P$ to a rumor set $R$ in $\Wu$: 
\begin{equation}
{\I_{\Wu}(P|R)} = \left\{ {\begin{array}{*{20}{l}}
	{\frac{1}{1+exp\{\alpha-\beta\cdot\ic_{\Wu}(P|R)\}}}	&	{\textrm{if $\ic_{\Wu}(P|R) > 0$}}\\
	0			&	\textrm{otherwise}
	\end{array}} \right.
\end{equation}
Here $\alpha$ and $\beta$ are the parameters that control the turning point of the user $u$ for being influenced by the protect information, where $\alpha$ controls the overall effectiveness of the influence block of $P$ to $R$ and $\beta$ controls the incremental effectiveness of influence block of one node in $P$ to $R$ in $\Wu$.
Then, let $ \I_{u}(\pr|\ru) = E[\I_{\Wu}(\pr|\ru)]$ for any $\wk_u$ denote the expected value of possibility that $P$ blocks the influence of $R$ to $u$.

\subsection{Problem definition}\label{definition}
Based on $ \I_{u}(\pr|\ru)$, the problem of Rumor Control when Impression Counts (\prob) can be described as follows.

\begin{defn}[Problem Definition]
	Given a graph $G = (V,E)$, an initial set $\ru \subset V$ and a budget $k$, $\prob$ is dedicated to finding a $k$-size set $\pr \subset V\backslash \ru$, which can maximize the influence block $\object({\pr}|{\ru}) = \sum\nolimits_{u \in V\backslash {\ru}} {{\I_u}} ({\pr}|{\ru})$.
\end{defn}

Next, we analyze the monotonicity and submodularity of $\object({\pr}|{\ru}) $ and the hardness of \prob.
%Given two sets of billboards S1 and S2, the marginal influence of adding S2 into S1 is Δ(S2 |S1) = I (S1 ∪ S2)−I (S1). Then, we define the monotonicity and submodularity of an influence function as follows. I (S) is monotone iff, I (S1) ≤ I (S2) for all S1 ⊆ S2. Furthermore, I (S) is submodular iff, given any set of billboards S∗, it satisfies Δ(S∗ |S1) ≥ Δ(S∗ |S2) for all S1 ⊆ S2. 
%The following presents a counterexample for the objective function to be submodular.

\begin{defn}
	We say that $\object({\pr}|{\ru}) $ is monotone iff, for any two assignment protector sets $P^a$ and $P^b$ such that $P^a\subseteq P^b$, it holds that $\object({\pr}^a|{\ru})\le\object({\pr}^b|{\ru})$. We say that $\object({\pr}|{\ru}) $ is submodular iff, for any two such protector sets and any  $P$, it has $\object({\pr}^a\cup P|{\ru}) - \object({\pr}^a|{\ru})\ge\object({\pr}^b\cup P|{\ru})-\object({\pr}^b|{\ru})$.
\end{defn}

It is trivial to show that $\object({\pr}|{\ru}) $ is monotone. However, as the following counterexample shows, $\object({\pr}|{\ru}) $ is not submodular.

\begin{figure}[t]
		\centering
		{\includegraphics[width=0.4\textwidth]{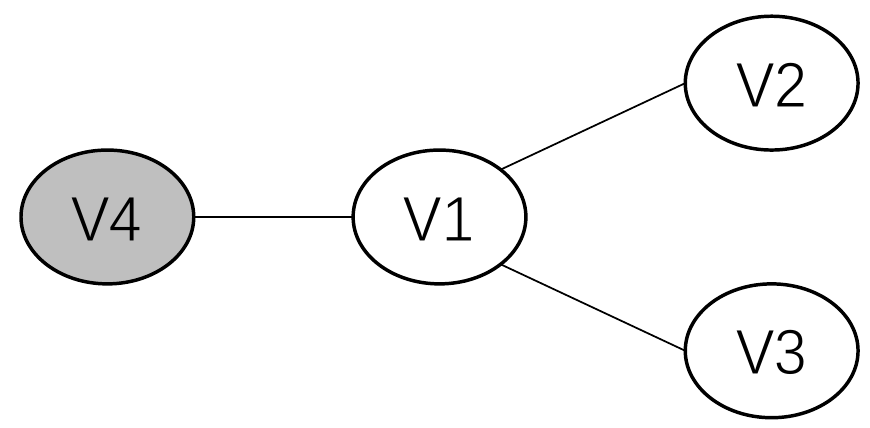}}
		\caption{An example of RCIC}
		\label{fig:example}
\end{figure}

\begin{figure}[t]
	
		\centering
		{\includegraphics[width=0.4\textwidth]{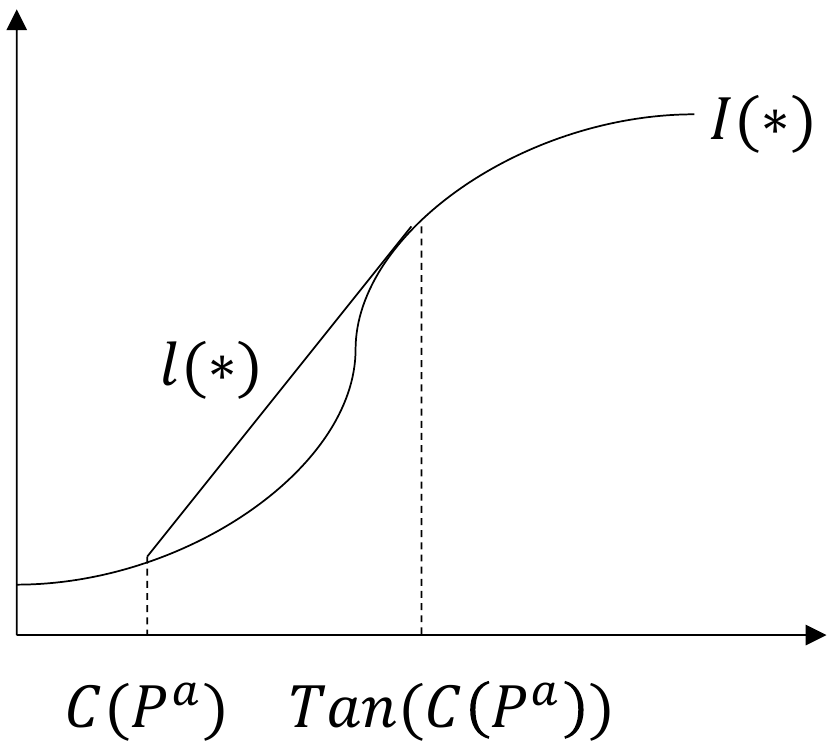}}
		\caption{The upper-bound influence block function}
		\label{fig:func}
\end{figure}

\begin{example}\label{non-sub}
	As shown in Figure~\ref{fig:example}, the rumor set $R=\{v4\}$. We choose $P^a=\{\}$, $P^b=\{v1\}$, $P=\{v2\}$, $\alpha = 3$, $\beta = 1$ and $T=2$. Then we have $\object(P^a|{\ru})=0$, $\object(P^b|{\ru})=1.372$ and $\object(P|{\ru})=0.358$. Furthermore, we have $\object({\pr}^a\cup P|{\ru}) - \object({\pr}^a|{\ru})=0.358$ and $\object({\pr}^b\cup P|{\ru})-\object({\pr}^b|{\ru}) = 2.433 -  1.372 = 1.061$. Since $P^a\subseteq P^b$ and $\object({\pr}^a\cup P|{\ru}) - \object({\pr}^a|{\ru})\le\object({\pr}^b\cup P|{\ru})-\object({\pr}^b|{\ru})$. We thus conclude $\object({\pr}|{\ru}) $ is not submodular.
\end{example}

\begin{thm}\label{NP-hardness}
	The $\prob$ problem is NP-hard.
\end{thm}
$Proof.$ We prove it by reducing the Set Cover problem to the $\prob$ problem. In the Set Cover problem, given a collection of subsets $S_1, S_2, ..., S_n$ of a universe of elements $U$ = $\{u_1,u_2,...,u_j\}$, we wish to know whether there exist $k$ of the subsets whose union is equal to $U$.
% %Consider an instance of the NP-complete Set Cover problem, defined by a collection of subsets $S_1, S2, ..., S_n$ of a ground set $U$ = $\{u_1,u_2,...,u_j\}$; we wish to know whether there exist $k$ of the subsets whose union is equal to $U$. (We can assume that $k < n$.) We show that this can be viewed as a special case of the weighted random walk domination problem.
We define a corresponding graph $G(V,E)$ with $n$ nodes. Each node in graph $G$ has $d$ edges connected. We set $\alpha$ as 0 and $\beta$ as $\infty$. Then we have ${\I_{\Wu}(P|R)} = 1$, if $\ic_{\Wu}(P|R) > 0$.

Given a rumor set $R$, we map a subset $S_i$ to a node $i$ in $V\backslash R$. Next, we generate all possible random walk instances with a total length equal to $T$ as a universe of elements $|U_0|$. Intuitively, $|U_0|=|V\backslash R|*d^T$. Then we map an element $u_j$ to a random walk instance $j$ in $U_0$ and $S_i$ contains $u_j$ when the random walk instance $j$ visits the node $i$ before visiting rumor set $R$. The Set Cover problem is equivalent to deciding whether there is a set $S$ of $k$ nodes in graph $G$ with $\object({\pr}|{\ru}) = |V\backslash R|$. As the set cover problem is NP-complete, the decision problem of \prob~is NP-complete, and the optimization problem is NP-hard. $\blacksquare$

%% file: tex/Ourframework.tex
\vspace{-0.3cm}\section{Our framework}
In this section, we first present a Monte Carlo-based greedy method (\samgreedy) as a baseline. Unfortunately, the effectiveness of this method is poor, and \samgreedy~cannot obtain any theoretical guarantees because the objective function of \prob~is non-submodular. Then we devise a Branch-and-Bound framework to solve this problem effectively. The core of this framework is how to estimate the upper bound of each candidate solution. In particular, we propose sampling-based bound estimation techniques for each branch under exploration by setting a submodular function to a tight upper bound of ${\I_{\Wu}(P|R)}$.

\subsection{A Baseline}
The core idea of \samgreedy~is to select the node $u$ which maximizes the unit marginal gain, i.e., $(\object({\pr} \cup \{ u \}|{\ru}) - \object({\pr}|{\ru}))$ , to a candidate solution set $\pr$, until the budget $k$ is exhausted. The pseudo-code of \samgreedy~is presented in Algorithm~\ref{al_samgreedy}. It first initializes $P$ as an empty set and $V\gets V \backslash R$. Next, it finds a set $P$ according to the greedy heuristic (Lines 1.6 to 1.10). In the end, it outputs set $P$ as a result.
\input{tex/al_samGreedy}
\subsection{Branch-and-Bound Framework}

As we analyzed above, \samgreedy~cannot obtain any theoretical guarantees because the objective function of \prob~is non-submodular. Then inspired by~\cite{DBLP:conf/kdd/ZhangLBMZ19}, we introduce a branch and bound framework to solve this problem effectively, and this solution can achieve a theoretical guarantee.

Algorithm~\ref{al_bab} shows the pseudo-code of the branch-and-bound framework.
We first initialize the global upper bound $U_G$ and global lower
bound $L_G$, and a max heap $H$ with each entry denoted as $\{P', V, U\}$, where $P'$ is the current node set that has been selected
as a protector set, $V$ is the set of a node that has not been considered yet, and $U$ is the upper bound influence block of the corresponding search space. $H$ is ordered by the upper bound value of each $P'$. While $L_G < U_G$, $H$ will pop the top entry that has the
maximum upper bound influence block. For each entry, if it matches the budget $k$ constraint, it will generate two new candidate sets ($P^a$ and $P^b$) by adding a new node $u \in V$ or not.
Then it computes the upper bound for each candidate set and updates $L_G$, $P$, and $H$ when $L^a>L_G$ and $U^a>L_G$, respectively.

\input{tex/al_BAB}
\subsection{Computing Upper Bound}

To estimate the upper bound of the current protector set $P^a$, we devise a submodular function ($\overline{\I}_{\Wu}(P|R)$ and $P = P^a \cup P^*$) as shown in Figure~\ref{fig:func} to compute the upper bound of $P^a$:\vspace{-0.2cm}
\begin{equation}
{\overline{\I}_{\Wu}(P|R)} = \left\{ {\begin{array}{*{20}{l}}
	l(\ic(P))	&	{\textrm{if $l(x)$ exists and }} \\
	&{ \textrm{$\ic(P^a) <\ic(P) < Tan(\ic(P^a))$}}\\
	\I_{\Wu}(P|R)			&	\textrm{otherwise}
	\end{array}} \right.\vspace{-0.2cm}
\end{equation}
Here, $\ic(P) = \ic_{\Wu}(P|R)$ for simplicity. $l(x)$ is the tangent through point ($\ic(P^a)$, $\I_{\Wu}(P^a|R)$) to function $\I_{\Wu}(P|R)$ and $Tan(\ic(P^a))$ is the x-coordinate of the tangent point. It is easy to see that $\overline{\I}_{\Wu}(P|R)$ is submodular as it concatenates two submodular functions for different domains. 

Furthermore, we have the following submodular function $\overline{\object}({\pr}|{\ru}) = \sum\nolimits_{u \in V\backslash {\ru}} {{\overline{\I}_u}} ({\pr}|{\ru})$ (here ${{\overline{\I}_u}} ({\pr}|{\ru}=  E[\overline{\I}_{\Wu}(\pr|\ru)]$ for any $\wk_u$) that
upper bounds the influence block function $\overline{\object}({\pr}|{\ru})$. It is also easy to see that $\overline{\object}({\pr}|{\ru}) $ is submodular as it is a sum of submodular functions. 

Due to the submodularity of $\overline{\object}({\pr}|{\ru})$, we turn to devise a greedy-based heuristic algorithm to find the upper bound for a given protector set $P^a$. In particular, we propose a sampling-based upper bound estimation algorithm to compute the upper bound for a given protector set.

\subsubsection{Sampling-based ComputeBound}
\input{./tex/al_GreedyBound}
As shown in algorithm~\ref{al_sbound}, it selects the node $u$ which maximizes the unit marginal gain%i.e., $(\overline{\object}({\pr} \cup \{ u \}|{\ru}) - \overline{\object}({\pr}|{\ru}))$ , 
to a candidate solution set $\pr$, until the budget $k$ is exhausted. 
%It is worth noting that we compute the unit marginal gain by the sampling random works.
In the end, it outputs set $P$, ${\object}(P|R)$ as $L^a$ and $\overline{\object}(P|R)$ as $U^a$.

%In order to obtain a more accurate approximation ratio, we devise a matrix-based solution for computing the upper bound in the following.

%\subsection{Matrix-based ComputeBound}
%\input{./tex/al_MatrixBound}

\subsection{Analysis of Solutions}\label{sec:Analysis}

In this section, we first analyze the approximate marginal gain computation in Algorithm~\ref{al_sbound} and show the proposed branch and bound framework with sampling-based computeBound can achieve a ($1-1/e-\epsilon$)-approximation factor through setting an appropriate sampling time $X$.
\vspace{-0.3cm}
\subsubsection{Approximate ratio of  SamComputeBound}\vspace{-0.2cm}

In algorithm~\ref{al_sbound}, it uses ${\overline{\I}'_u} ({\pr}|{\ru})$ which is computed according to the random walk sampling set as an estimator of ${\overline{\I}_u}({\pr}|{\ru})$. To estimate the expectation of ${\overline{\I}_{\Wu}(P|R)}$, we independently run $X$ random walks starting from $u$, and take the average of $\overline{\I}_{\Wu}(P|R) $ as the estimator. The proposed sampling process is equivalent to a simple random sampling with replacement, thus the estimator is unbiased. Then we use $\overline{\object}'({\pr}|{\ru}) = \sum\nolimits_{u \in V\backslash {\ru}} {{\overline{\I}'_u}} ({\pr}|{\ru})$ as a estimator of $\overline{\object}({\pr}|{\ru})$.

Next, we apply Hoeffding’s inequality~\cite{hoeffding1994probability} to bound the sample size $X$. Specifically, we have the following lemma.

\begin{lem}\label{Hoeffding}
	Given a protector set $P$ and a rumor set $R$, for two small constants $\epsilon$ and $\delta$, if $X \ge \frac{1}{2\epsilon^2}log\frac{n -|R|}{\delta}$, then $\mathbf{Pr}[|\overline{\object}'({\pr}|{\ru})  - \overline{\object}({\pr}|{\ru})  | \ge\epsilon(n -|R||)] \le \delta$.
\end{lem}

$Proof.$ First, we have %$\mathbf{Pr}[|\overline{\object}'({\pr}|{\ru})  - \overline{\object}({\pr}|{\ru})  | \ge \epsilon(n -|R|)]\le \mathbf{Pr}[\sum\nolimits_{u \in V\backslash {\ru}} |{{\overline{\I}'_u}} ({\pr}|{\ru}) - {{\overline{\I}_u}} ({\pr}|{\ru})| \ge \epsilon(n -|R|)]$,
\begin{equation*}
\begin{aligned}
\mathbf{Pr}[&|\overline{\object}'({\pr}|{\ru})  - \overline{\object}({\pr}|{\ru})  | \ge \epsilon(n -|R|)]\\&\le \mathbf{Pr}[\sum\nolimits_{u \in V\backslash {\ru}} |{{\overline{\I}'_u}} ({\pr}|{\ru}) - {{\overline{\I}_u}} ({\pr}|{\ru})| \ge \epsilon(n -|R|)],
\end{aligned}
\end{equation*}
as $|\overline{\object}'({\pr}|{\ru})  - \overline{\object}({\pr}|{\ru})  | \ge\epsilon(n -|R|) \ge \epsilon(n -|R|)$ implies $\sum\nolimits_{u \in V\backslash {\ru}} |{{\overline{\I}'_u}} ({\pr}|{\ru}) - {{\overline{\I}_u}} ({\pr}|{\ru})|\ge \epsilon(n -|R|)$. Then, by the union bound, we have %$\mathbf{Pr}[\sum\nolimits_{u \in V\backslash {\ru}} |{{\overline{\I}'_u}} ({\pr}|{\ru}) - {{\overline{\I}_u}} ({\pr}|{\ru})| \ge \epsilon(n -|R|)] \le  \sum\nolimits_{u \in V\backslash {\ru}} \mathbf{Pr}[|({{\overline{\I}'_u}} ({\pr}|{\ru}) - {{\overline{\I}_u}} ({\pr}|{\ru})) \ge \epsilon].$
\begin{equation*}
\begin{aligned}
\mathbf{Pr}[\sum\nolimits_{u \in V\backslash {\ru}} &|{{\overline{\I}'_u}} ({\pr}|{\ru}) - {{\overline{\I}_u}} ({\pr}|{\ru})| \ge \epsilon(n -|R|)] \le \\ &\sum\nolimits_{u \in V\backslash {\ru}} \mathbf{Pr}[|({{\overline{\I}'_u}} ({\pr}|{\ru}) - {{\overline{\I}_u}} ({\pr}|{\ru})) \ge \epsilon].
\end{aligned}
\end{equation*}

Since $0 \le {{\overline{\I}_u}} ({\pr}|{\ru})) \le 1$, we can apply Hoeffding’s
inequality~\cite{hoeffding1994probability} to bound the sample size $X$. Specifically, we
have %$\mathbf{Pr}[|({{\overline{\I}'_u}} ({\pr}|{\ru}) - {{\overline{\I}_u}} ({\pr}|{\ru}))| \ge \epsilon]\le exp(-2\epsilon^2X)$.
\begin{equation*}
\begin{aligned}
\mathbf{Pr}[|({{\overline{\I}'_u}} ({\pr}|{\ru}) - {{\overline{\I}_u}} ({\pr}|{\ru}))| \ge \epsilon]\le exp(-2\epsilon^2X).
\end{aligned}
\end{equation*}
Based on this, the following inequality immediately holds %$\mathbf{Pr}[|\overline{\object}'({\pr}|{\ru})  - \overline{\object}({\pr}|{\ru})  | \ge\epsilon(n -|R|)] \le (n-|R|)exp(-2\epsilon^2X)$
\begin{equation*}
\begin{aligned}
\mathbf{Pr}[|&\overline{\object}'({\pr}|{\ru})  - \overline{\object}({\pr}|{\ru})  | \\&\ge\epsilon(n -|R|)] \le (n-|R|)exp(-2\epsilon^2X).
\end{aligned}
\end{equation*}
Let $(n-|R|)exp(-2\epsilon^2X) \le \delta$, then we can get $X \ge \frac{1}{2\epsilon^2}log\frac{n -|R|}{\delta}$, which completes the proof. $\blacksquare$

According to \cite{nemhauser1978analysis}, the greedy heuristic achieves an approximation factor
of $(1 - 1/e)$ for maximizing monotone and submodular functions. Based on Lemma~\ref{Hoeffding}, by a similar analysis presented in \cite{DBLP:conf/icde/LiYHC14}, the sampling-based greedy algorithm achieves a ($1-1/e-\epsilon$) approximation factor through setting an appropriate parameter $X$ with at least ($1-\delta$) probability.

\subsubsection{Approximate ratio of branch and bound}
The upper bounding techniques lead to a constant approximation ratio for the solution returned by the branch and bound framework. In particular, we have the following theorem.

\begin{thm}\label{ratio}
	The branch and bound framework with sampling-based computeBound achieves an approximation factor of ($1-1/e-\epsilon$) for the $\prob$ through setting an appropriate parameter $X$.
\end{thm}
$Proof.$ Let $P$ denote the solution outputted by Algorithm~3 and $P^*$ denote the optimal solution for sampling-based computeBound.
As we analyzed above, Algorithm~3 achieves a ($1-1/e-\epsilon$) approximation factor through setting an appropriate parameter $X$ with at least ($1-\delta$) probability. Then we have %$\overline{\object}(P|{\ru}) \ge (1-1/e-\epsilon)(\overline{\object}(P^*|{\ru})\ge (1-1/e-\epsilon)({\object}(P^* |{\ru}).$
\begin{equation*}
\begin{aligned}
\overline{\object}(P|{\ru}) &\ge (1-1/e-\epsilon)(\overline{\object}(P^*|{\ru})\\&\ge (1-1/e-\epsilon)({\object}(P^* |{\ru}).
\end{aligned}
\end{equation*}

Let $P_{out}$ denote the returned solution by Algorithm~2. For any branch that has not been searched, under the termination condition $L<U$. Then we have $\object(P_{out}|{\ru})\ge \overline{\object}(P|{\ru})$. Therefore, Algorithm~2 achieves $\object(P_{out}|{\ru})\ge (1-1/e-\epsilon)({\object}(P^* |{\ru})$.
$\blacksquare$

%% file: tex/al_samGreedy.tex
%\vspace{-0.5cm}
%\setlength{\algomargin}{0.5em} 
%\setlength{\belowcaptionskip}{-3cm}
%\setlength{\intextsep}{0.5cm}
\setlength{\textfloatsep}{0pt}
\begin{algorithm}[t]
\caption{\samgreedy$(G, R,, k)$} 
\label{al_samgreedy}
\DontPrintSemicolon
\begin{small}
{\bf Input:} a graph $G$, a rumor set $R$%, sampling time $X$ for each node 
and a budget $k$

{\bf Output:} a protector set $P$

Run $X$ random walks for each node in $V\backslash R$;
$\IS(\ru) \gets$ all the random walks influenced by $R$;
Initialize $P$ as an empty set and $V \gets V \backslash R$.

\Repeat{$k = 0$}
{
	%\tcc{$\object(P|R)$ is computed based on Monte Carlo simulations $\IS(\ru)$.}
	
	Select $u \gets \arg \max _{v \in V}((\object({\pr} \cup \{ v \}|{\ru}) - \object({\pr}|{\ru})))$
	
	$V \gets V \backslash \{u\}$ and $P \gets P \cup \{u\}$;
	$k \gets k - 1$
	
}

\Return {$P$}
\end{small}
\end{algorithm}

%% file: tex/al_BAB.tex
\vspace{-0.5cm}
\begin{algorithm}[t]
\caption{\bab$(G, R, k)$} 

\label{al_bab}
\DontPrintSemicolon
\begin{small}
{\bf Input:} a graph $G$, a rumor set $R$ and a budget $k$

{\bf Output:} a protector set $P$

Initialize $P$ and $P'$ as an empty set and $V \gets V \backslash R$;
$L_G \gets 0$ and $U_G \gets \infty$;
Initialize max heap $H \gets \{P', V, U\}$

\Repeat{$L_G\ge U_G$}
{
	$\{P', V, U\} \gets$ top of $H$;
	Select $u \in V$
	
	\If{$|P'| < k$}
	{
		$V \gets V \backslash u$
		
		$P^a \gets P'\cup u$ and  $P^b \gets P'$
		
		$ \{P^c, L^a, U^a\} \gets \sbound(P^a, V)$ \label{bab:start}
		
		\If{$ L^a > L_G$}
		{
			$L_G\gets L^a$ and $P\gets P^c$
		}
		
		\If{$ U^a > L_G$}
		{
			$H\gets H\cup \{P^a, V, U^a\} $
		}\label{bab:end}
		
		Repeat line~\ref{bab:start} to \ref{bab:end} for $P^b$
	}
	
}

\Return {$P$}
\end{small}
\end{algorithm}

%% file: tex/al_GreedyBound.tex
\begin{algorithm}[t]
\caption{\sbound$(P^a, V)$} 

\label{al_sbound}
\DontPrintSemicolon
\begin{small}
{\bf Input:} Protector set $P^a$ and candidate node set $V$

{\bf Output:} $ \{P, L^a, U^a\}$

Run $X$ $T$-random walks for each node in $V$;
$\IS(\ru) \gets$ all the random walks influenced by $R$;
Initialize $P$ as $P^a$ and $k\gets k-|P^a|$

\Repeat{$k = 0$}
{
	%\tcc{$\overline{\object}(P|R)$ is computed based on Monte Carlo simulations $\IS(\ru)$.}
	
	Select $u \gets \arg \max _{v \in V}((\overline{\object}({\pr} \cup \{ v \}|{\ru}) - \overline{\object}({\pr}|{\ru})))$
	
	$V \gets V \backslash \{u\}$ and $P \gets P \cup \{u\}$;
	$k \gets k - 1$
	
}

\Return {$P$, $L^a\gets {\object}(P|R)$, $U^a\gets \overline{\object}(P|R)$}
\end{small}
\end{algorithm}

%% file: tex/pro.tex
\section{Progressive Branch-and-Bound}
Although  Algorithm~\ref{al_bab}~improves the effectiveness of basic greedy by conducting the branch-and-bound framework, it still suffers from a high computational cost due to heavily invoking Algorithm~\ref{al_sbound} for bound estimations. To be more specific, in each greedy search iteration of Algorithm~\ref{al_sbound}, it has to recalculate the marginal gain $(\overline{\object}({\pr} \cup \{ v \}|{\ru}) - \overline{\object}({\pr}|{\ru}))$ for all candidate nodes.

Motivated by this observation, we propose a progressive  sampling-based upper bound estimation method (\psbound). It selects multiple, but not only one, nodes in each greedy search iteration to cut down the total number of iterations required and hence the computation cost. Meanwhile, we will prove that it can achieve an approximation ratio of $(1 - 1/e - \epsilon - \rho)$ for the upper bound estimation, where $\rho$ is a tunable parameter that provides a trade-off between efficiency and accuracy.

The pseudo-code of \psbound~is shown in Algorithm~\ref{al_psbound}. \psbound~first sorts $v \in V$ based on descending order of $\overline{\object}_{v}(P|R)$ and initializes the threshold h to the value of $\max _{v \in V}\overline{\object}_{v}(P|R)$. Then, it iteratively fetches all the nodes with their marginal gains not smaller than $h$ into $P$ and meanwhile lowers the threshold $h$ by a factor of $(1 + \rho)$ for the next iteration (Lines~\ref{pro:start}-\ref{pro:end}). The iteration continues until there are $k$ nodes in $P$. Unlike the basic greedy method that has to check all the potential nodes in candidate node set $V$ in each iteration, it is not necessary for \psbound~as it implements an early termination (Lines~\ref{pro:s1}-\ref{pro:e1}). Since nodes are sorted by $\overline{\object}_{v}(P|R)$ values, if $\overline{\object}_{v}(P|R)$ of the current node is smaller than $h$, all the nodes $v‘$ pending for evaluation will have their $\overline{\object}_{v‘}(P|R)$ values smaller than $h$ and hence could be skipped from evaluation. 

In the following, we first analyze the approximation ratio of Algorithm~\ref{al_psbound}  for upper bound estimation by Lemma 2. Based on Lemma 2, we show the approximation ratio of the branch-and-bound framework invoking Algorithm~\ref{al_psbound} for \prob~by Theorem 3.

\begin{lem}\label{ratio:psbound}
	\psbound~achieves a $(1 - 1/e - \epsilon - \rho)$ approximation ratio for upper bound estimation.
\end{lem}
$Proof.$ We first prove \psbound~achieves a $(1 - 1/e - \rho)$ approximation ratio for maximizing monotone and submodular functions. At this stage, we do not consider estimating the marginal gains based on the sampling results but assume that the true marginal gains can be obtained. Then we show \psbound~achieves a $(1 - 1/e - \epsilon - \rho)$ approximation ratio for upper bound estimation. 

\noindent
\textbf{\psbound~for maximizing monotone and submodular functions.}  For a given rumor set $R$, let $v_i$ be the node selected at a given threshold $h$ and $O$ denote the optimal
local solution to the problem of selecting $k$ nodes that can maximize $\overline{\object}$. Because of the submodularity of $\overline{\object}$, we have
\begin{equation} \label{appro1}
\overline{\object}_{v}(P|R)= \left\{ \begin{array}{l}
\ge h\\
\le h \cdot (1 + \rho ) 
\end{array} \right. \begin{array}{*{20}{l}}
{ \textrm{if $v=v_i$}}\\
{\textrm{if $v\in O\backslash(P\cup v_i)$}},
\end{array}
\end{equation}
where $P$ is the current partial solution. Equation~(\ref{appro1}) implies that $\overline{\object}_{v_i}(P|R)\geq\overline{\object}_{v}(P|R)/(1+\varepsilon)$ for any $v\in O\backslash P$. Thus, we have %$\overline{\object}_{v_i}(P|R) \geq \frac{1}{(1+\rho)|\OPT\backslash P|}\sum\nolimits_{v \in \OPT\backslash P}\overline{\object}_{v}(P|R)\geq \frac{1}{(1+\varepsilon)n}\sum\nolimits_{v \in \OPT\backslash P}\overline{\object}_{v}(P|R)$.
\begin{equation*}
\begin{aligned}
	\overline{\object}_{v_i}(P|R) &\geq \frac{1}{(1+\rho)|\OPT\backslash P|}\sum\nolimits_{v \in \OPT\backslash P}\overline{\object}_{v}(P|R)\\
	&\geq \frac{1}{(1+\varepsilon)n}\sum\nolimits_{v \in \OPT\backslash P}\overline{\object}_{v}(P|R).
\end{aligned}
\end{equation*}
Let $P_i$ denote the partial solution that $v_i$ has been included and $v_{i+1}$ be the node selected at the $(i+1)$th step. Then we have %$\overline{\object}(P_{i+1}|R)-\overline{\object}(P_i|R)=\overline{\object}_{v_i}(P_i|R) \geq \frac{1}{(1+\rho)n}\sum\nolimits_{v \in \OPT\backslash P_i}\overline{\object}_{v}(P_i|R) \geq \frac{1}{(1+\rho)n}(\overline{\object}(\OPT\cup P_{i}|R)-\overline{\object}(P_{i}|R))\geq \frac{1}{(1+\rho)n}(\overline{\object}(\OPT|R)-\overline{\object}(P_{i}|R))$. 
\begin{equation*}
	\begin{aligned}
	\overline{\object}(P_{i+1}|R)&-\overline{\object}(P_i|R)=\overline{\object}_{v_i}(P_i|R)\\
	&\geq \frac{1}{(1+\rho)n}\sum\nolimits_{v \in \OPT\backslash P_i}\overline{\object}_{v}(P_i|R)\\
	&\geq \frac{1}{(1+\rho)n}(\overline{\object}(\OPT\cup P_{i}|R)-\overline{\object}(P_{i}|R))\\
	&\geq \frac{1}{(1+\rho)n}(\overline{\object}(\OPT|R)-\overline{\object}(P_{i}|R)).
	\end{aligned}
\end{equation*}

The solution $P^*$ obtained by Algorithm~4 with $|P^*|= k$. Using the geometric series formula, we have %$\overline{\object}(P^{*}|R)\ge\left( {1 - \left( 1- \frac{{1}}{{(1 + \rho )n}}\right)^n} \right)\overline{\object}\left( \OPT |R\right)\ge\left( {1 - {e^{\frac{{ - n}}{{(1 + \rho )n}}}}} \right)\overline{\object}\left( \OPT |R\right)=\left( {1 - {e^{\frac{{ - 1}}{{(1 + \rho )}}}}} \right)\overline{\object}\left( \OPT |R\right)\ge\left( {(1-1/e-\rho)} \right)\overline{\object}\left( \OPT |R\right)$. 
\begin{equation*}
	\begin{aligned}
	\overline{\object}(P^{*}|R)&\ge\left( {1 - \left( 1- \frac{{1}}{{(1 + \rho )n}}\right)^n} \right)\overline{\object}\left( \OPT |R\right)\\
	&\ge\left( {1 - {e^{\frac{{ - n}}{{(1 + \rho )n}}}}} \right)\overline{\object}\left( \OPT |R\right)\\
	&=\left( {1 - {e^{\frac{{ - 1}}{{(1 + \rho )}}}}} \right)\overline{\object}\left( \OPT |R\right)\\
	&\ge\left( {(1-1/e-\rho)} \right)\overline{\object}\left( \OPT |R\right).
	\end{aligned}
\end{equation*}

Hence, that \psbound~achieves a $(1 - 1/e - \rho)$ approximation ratio for maximizing monotone and submodular functions has been proved.

\noindent
\textbf{\psbound~for upper bound estimation.}  As we analyzed above, Algorithm 4  achieves an approximation factor
of $(1 - 1/e - \rho)$ for maximizing monotone and submodular functions. Based on Lemma~\ref{Hoeffding}, by a similar analysis presented in \cite{DBLP:conf/icde/LiYHC14}, the progressive sampling-based greedy algorithm achieves a ($1-1/e-\epsilon - \rho$) approximation factor through setting an appropriate parameter $X$ with at least ($1-\delta$) probability for upper bound estimation. $\blacksquare$

\begin{thm}\label{proratio}
	The branch and bound framework with sampling-based computeBound achieves an approximation factor of ($1-1/e-\epsilon- \rho$) for the $\prob$ through setting an appropriate parameter $X$.
\end{thm}
$Proof.$ Similar to the proof of Theorem~\ref{ratio}.
Let $P$ denote the solution outputted by Algorithm~4 and $P^*$ denote the optimal solution for progressive sampling-based computeBound.
As we analyzed above, Algorithm~4 achieves a ($1-1/e-\epsilon - \rho$) approximation factor by setting an appropriate parameter $X$ with at least ($1-\delta$) probability. Then we have 
\begin{equation*}
\begin{aligned}
\overline{\object}(P|{\ru}) &\ge (1-1/e-\epsilon -\rho)(\overline{\object}(P^*|{\ru})\\&\ge (1-1/e-\epsilon -\rho)({\object}(P^* |{\ru}).
\end{aligned}
\end{equation*}
Let $P_{out}$ denote the returned solution by Algorithm~2. For any branch that has not been searched, under the termination condition $L<U$. Then we have $\object(P_{out}|{\ru})\ge \overline{\object}(P|{\ru})$. Therefore, Algorithm~2 achieves $\object(P_{out}|{\ru})\ge (1-1/e-\epsilon  - \rho)({\object}(P^* |{\ru})$.
$\blacksquare$

\input{./tex/al_ProGreedyBound}

%% file: tex/al_ProGreedyBound.tex
\begin{algorithm}[t]
\caption{\psbound$(P^a, V)$} 

\label{al_psbound}
\DontPrintSemicolon
\begin{small}
{\bf Input:} Protector set $P^a$ and candidate node set $V$

{\bf Output:} $ \{P, L^a, U^a\}$

Run $X$ $T$-random walks for each node in $V$;
$\IS(\ru) \gets$ all the random walks influenced by $R$;
Initialize $P$ as $P^a$;
Sort $v \in V$ based on descending order of $\overline{\object}_{v}(P|R)$;
Initialize $h \gets \max _{v \in V}\overline{\object}_{v}(P|R)$

\While{$|P|\leq k$}
{
	\For{each $v \in V$}{ \label{pro:start}
		
		\If{$|P|\leq k$}{
			
			$\overline{\object}_{v}(P|R) \gets (\overline{\object}({\pr} \cup \{ v \}|{\ru}) - \overline{\object}({\pr}|{\ru}))$
			
			\If{$\overline{\object}_{v}(P|R)\geq h$}{
				$P\gets P\cup v$, $V \gets V \backslash v$
			}
			
			%\If{$\G_{b_{jl}}(\phi) < h$}{ \label{pro:s1}
			\If{$\overline{\object}_{v}(P|R)< h$}{ \label{pro:s1}
				\textbf{break}
			}\label{pro:e1}
		}\Else{
			\textbf{break}
		}
		
	}
	
	$h \gets  \frac{h}{1+\rho}$\label{pro:end}
	
}

\Return {$P$, $L^a\gets {\object}(P|R)$, $U^a\gets \overline{\object}(P|R)$}
\end{small}
\end{algorithm}

%% file: tex/exp.tex
\section{Experiments}\label{sec_5}
In this section, we present our experimental results on the effectiveness, efficiency, memory consumption, and scalability of our proposed methods.
\subsection{Experimental settings}
\noindent
\textbf{DataSets.} We use three real-world datasets in the experiments: Gnutella, Email-Enron, and Gowalla. All the datasets are obtained from an open-source website\footnote{http://snap.stanford.edu/data/}, and
their statistics are shown in Table~\ref{tab_datasets}. The Gnutella dataset is a peer-to-peer file-sharing network, the Email-Enron dataset is an email communication network, and the Gowalla dataset is a location-based social networking website where users share their locations by checking in.

\begin{table}[t]	
		\centering
	\caption{Parameter setting.}
	\begin{small}
		\label{exp_param}
		\begin{tabular}[r]{|p{1.8cm}<{\centering}|p{4cm}<{\centering}|}
			\hline
			\multicolumn{1}{|c|}{Parameters}                                  & \multicolumn{1}{c|}{Values} \\ \hline
			$k$                                                       &	50, 100, \textbf{150}, 200, 250\\ \hline
			$|R|$                                                             &	50, 100, \textbf{150}, 200, 250\\ \hline
			$T$                                                               &	3, 6, \textbf{9}, 12, 15    \\ \hline
			$\beta/\alpha$                                                               &	\textbf{3/7}, 3/8, 3/9, 3/10, 3/11    \\ \hline
			$X$                                                               &	500, \textbf{1000}, 1500, 2000, 2500    \\ \hline
			$\rho$                                                               &	0.0001, 0.001, 0.01, \textbf{0.1}, 1    \\ \hline
		\end{tabular}
	\end{small}
\end{table}

\begin{table}[t]
	\caption{Summary of the datasets.}
	\centering
	\begin{small}
		\label{tab_datasets}
		\begin{tabular}[c]{|c|c|c|c|c|}
			\hline
			&$n$			& $m$		     &\#AvgDegree       &\#MaxDegree 	 \\ \hline
			Gnutella    	&8.8k     	 	&63k      	 &7.2    		    &88   		\\ \hline
			Email-Enron    	&37k     		&184k       	 &5.01   		    &1383  		 	 \\ \hline
			Gowalla     	&197k     	&950k          &4.83      	    &14730   	   \\ \hline    	
		\end{tabular}
	\end{small}
\end{table}

\noindent
\textbf{Algorithms.} To the best of our knowledge, this is the first work to study \prob, and thus there exists no previous work for direct comparison. In particular, we compare the four following methods. (1) \degreeTop: It is to select the top-$k$ high block degree nodes in the sampling random walk set as the targeted nodes. (2) \samgreedy: A basic sampling-based greedy algorithm~(Algorithm \ref{al_samgreedy}). (3) \bab~(BAB): The branch-and-bound framework (Algorithm \ref{al_bab}) with Algorithm~\ref{al_sbound} for bound estimations. (4) Progressive \bab~(ProBAB): The branch-and-bound framework (Algorithm \ref{al_bab}) with Algorithm~\ref{al_psbound} for bound estimations.

% \begin{itemize}
% 	\item \degreeTop: It is to select the top-$k$ high block degree nodes in sampling random walk set as the targeted nodes.
% 	\item \samgreedy: A basic sampling-based greedy algorithm~(Algorithm \ref{al_samgreedy}). In each iteration, it selects the node $u$ which maximizes the marginal gain  $(\object({\pr} \cup \{ u \}|{\ru}) - \object({\pr}|{\ru}))$ to a candidate solution set $\pr$, until the budget $k$ is exhausted.
% 	\item \bab~(BAB): The branch-and-bound framework (Algorithm \ref{al_bab}) with Algorithm~\ref{al_sbound} for bound estimations.
% 	\item Progressive \bab~(ProBAB): The branch-and-bound framework (Algorithm \ref{al_bab}) with Algorithm~\ref{al_psbound} for bound estimations.
% \end{itemize}

%\degreeTop, \samgreedy~(Algorithm \ref{al_samgreedy}), and \bab~(BAB for short, Algorithm \ref{al_bab}). 

\noindent
\textbf{Evaluation metrics.} We evaluate the performance of all methods by the runtime and the blocking percentage of the selected nodes. In particular, the percentage is computed by $\object({\pr}|{\ru})/\IS(\ru)$, where $\IS(\ru)$ denote the random walk set influenced by rumor set $\ru$. %Moreover, to accurately measure $\object({\pr}|{\ru})$ for each algorithm, we compute it by running a sufficient number of MC simulations, i.e., $X = 1000$.

\noindent
\textbf{Parameter.} Table \ref{exp_param} shows the settings of all parameters, such as the budget $k$, the size of the rumor set $R$, the (random walk) length threshold $T$, the number of samples $X$, the parameter $\alpha$, the parameter $\beta$ and parameter $\rho$.
Here the default one is highlighted in bold.
To simulate the rumor set $R$, we select nodes uniformly at random from the nodes whose degrees are in the top 10\% of $G$.

% \begin{table}[t]
% 	\caption{Parameter setting.}
% 	\centering
% 	\begin{small}
% 		\label{exp_param}
% 		\begin{tabular}[r]{|p{1.8cm}<{\centering}|p{4cm}<{\centering}|}
% 			\hline
% 			\multicolumn{1}{|c|}{Parameters}                                  & \multicolumn{1}{c|}{Values} \\ \hline
% 			$k$                                                       &	50, 100, \textbf{150}, 200, 250\\ \hline
% 			$|R|$                                                             &	50, 100, \textbf{150}, 200, 250\\ \hline
% 			$T$                                                               &	3, 6, \textbf{9}, 12, 15    \\ \hline
% 			%$\beta/\alpha$                                                               &	\textbf{3/7}, 3/8, 3/9, 3/10, 3/11    \\ \hline
% 			%$X$                                                               &	500, \textbf{1000}, 1500, 2000, 2500    \\ \hline
% 			%$\rho$                                                               &	0.0001, 0.001, 0.01, \textbf{0.1}, 1    \\ \hline
% 		\end{tabular}
% 	\end{small}
% \end{table}

\noindent
\textbf{Setup.} All codes are implemented in Java, and experiments are conducted on a server with 2.1 GHz Intel Xeon 8 Core CPU and 32GB memory running CentOS/6.8 OS.
\input{./tex/Effectiveness}
\input{./tex/Efficiency}
\input{./tex/Memory}
\input{./tex/Scalability}

%% file: tex/Effectiveness.tex
\vspace{-0.3cm}\subsection{Effectiveness test}\label{sec_effectiveness}\vspace{-0.2cm}
This section studies how the block degree is affected by varying the budget $k$, the size of the rumor set $R$, and the length threshold $T$ of a random walk.

\noindent
\textbf{Varying the budget $k$.} The block degrees of all algorithms on Gnutella and Email-Enron by varying the $k$ are shown in Figure \ref{fig:exp_ef_Gnutella_B} and Figure \ref{fig:exp_ef_Email_B}, respectively, and we find that when the budget raises from 50 to 250,
BAB outperforms Greedy and TopK by up to 115\% in the Email-Enron.

\noindent
\textbf{Varying the size of $R$.} Figure \ref{fig:exp_ef_Gnutella_R} and Figure \ref{fig:exp_ef_Email_R}  show the result by varying the size of $R$. We find: (1) with the growth of $|R|$, the blocking percentages of all methods are increasing because the increasing influence of $R$ leads to more nodes with higher unit block degrees. (2) ProBAB and BAB are consistently better than that of the rest baselines.

\noindent
\textbf{Varying the random walk length threshold $T$.} Figure \ref{fig:exp_ef_Gnutella_T} and Figure \ref{fig:exp_ef_Email_T}  show the results by varying the threshold $T$, which determines the length of a random walk starting from a node. We observe that: (1) The rumors on Gnutella dataset are much harder to be controlled than Email-Enron dataset. It implies that the network structure is an important variable for \prob. (2) With the increase of $T$, the performance of all algorithms becomes better. The reason is that when the length becomes large, the random walk has more chances to reach the protectors and thus leads to a high unit block degree of the seeds.

\begin{figure*}[t]
	\centering
	%\includegraphics[width=1\textwidth]{exp/title5.pdf}
	%\vspace{-0.3cm}
	%\hspace{-10pt}
	\subfloat[Varying $k$]
	{\label{fig:exp_ef_Gnutella_B}\includegraphics[width=0.3\textwidth]{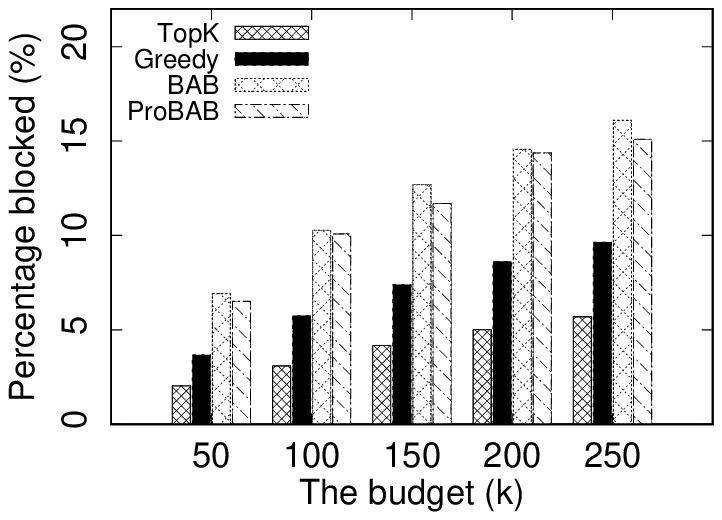}}
	%\hspace{-10pt}
	\subfloat[Varying $|R|$]
	{\label{fig:exp_ef_Gnutella_R}\includegraphics[width=0.3\textwidth]{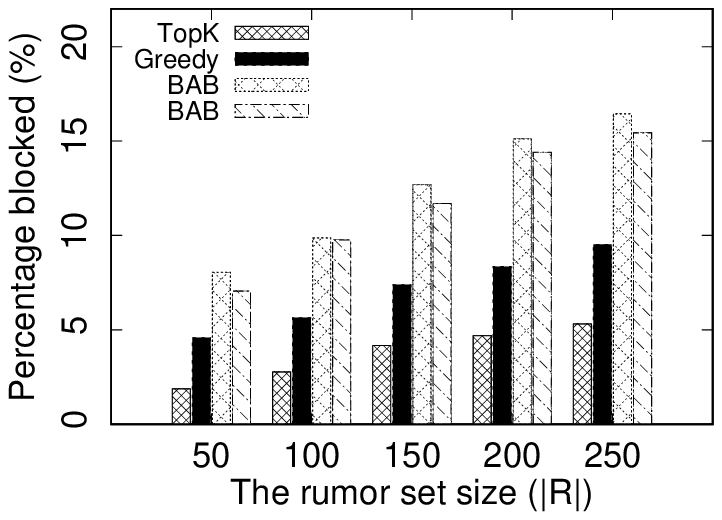}}
	%\hspace{-10pt}
	\subfloat[Varying $T$]
	{\label{fig:exp_ef_Gnutella_T}\includegraphics[width=0.3\textwidth]{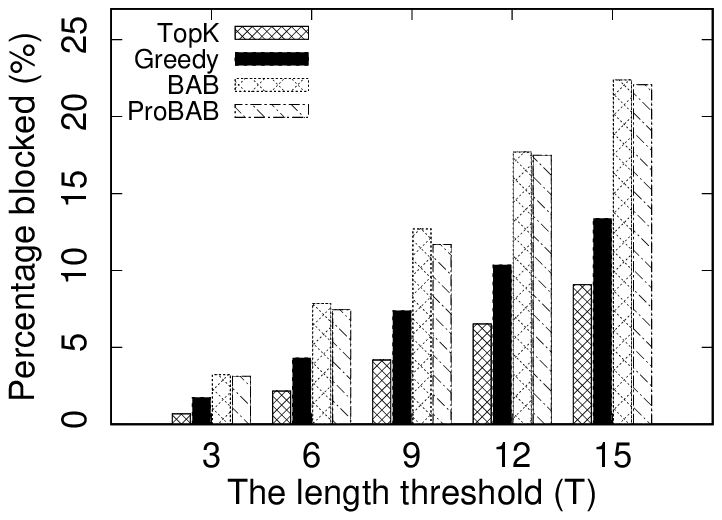}}
	\caption{Effectiveness test on Gnutella}
	\label{fig:para}
\end{figure*}
\begin{figure*}[t]
	\centering
	%\includegraphics[width=1\textwidth]{exp/title5.pdf}
	%\vspace{-0.3cm}
	%\hspace{-10pt}
	\subfloat[Varying $k$]
	{\label{fig:exp_ef_Email_B}\includegraphics[width=0.3\textwidth]{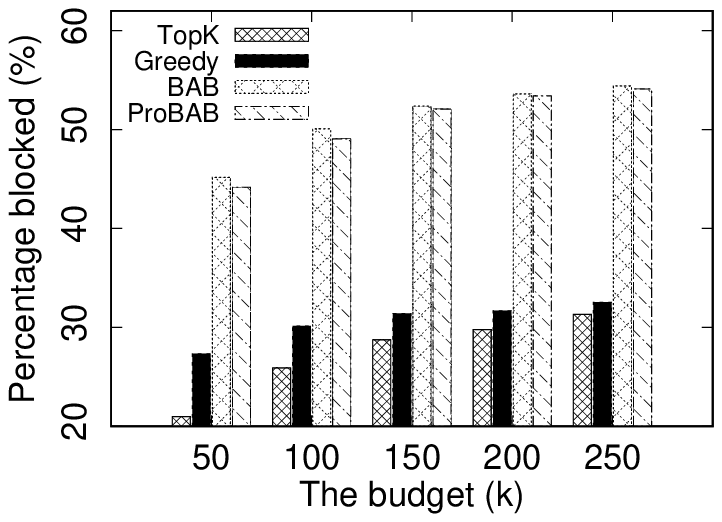}}
	%\hspace{-10pt}
	\subfloat[Varying $|R|$]
	{\label{fig:exp_ef_Email_R}\includegraphics[width=0.3\textwidth]{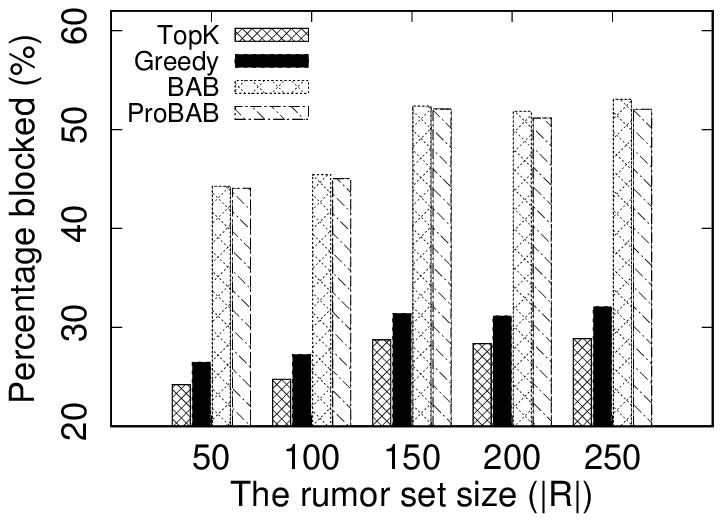}}
	%\hspace{-10pt}
	\subfloat[Varying $T$]
	{\label{fig:exp_ef_Email_T}\includegraphics[width=0.3\textwidth]{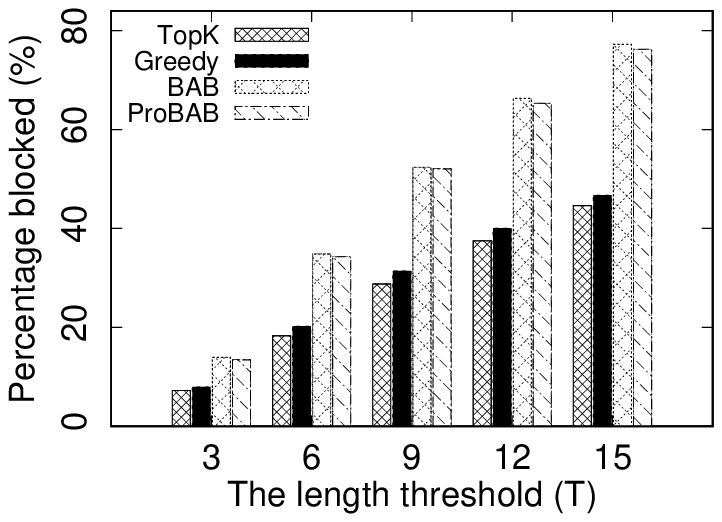}}
	\caption{Effectiveness test on Email-Enron}
	\label{fig:para}
\end{figure*}

\iffalse
\begin{figure}[t]
	\centering
	\subfloat[Gnutella]{\includegraphics[clip,width=0.245\textwidth]{exp/k-effect-G.eps}\label{fig:exp_ef_Gnutella_B}}
	\hspace{-5pt}
	\subfloat[Email-Enron]{\includegraphics[clip,width=0.245\textwidth]{exp/k-effect-E.eps}\label{fig:exp_ef_Email_B}}
	\caption{Effectiveness of varying the budget $k$}
	\label{fig:exp_ef_B}
\end{figure}
\fi

\iffalse
\begin{figure}[t]
	\centering
	\subfloat[Gnutella]{\includegraphics[clip,width=0.245\textwidth]{exp/R-effect-G.eps}\label{fig:exp_ef_Gnutella_R}}
	\hspace{-5pt}
	\subfloat[Email-Enron]{\includegraphics[clip,width=0.245\textwidth]{exp/R-effect-E.eps}\label{fig:exp_ef_Email_R}}
	\caption{Effectiveness of varying the size of $R$}
	\label{fig:exp_ef_R}
\end{figure}
\fi

\iffalse
\begin{figure}[t]
	\centering
	\subfloat[Gnutella]{\includegraphics[clip,width=0.245\textwidth]{exp/L-effect-G.eps}\label{fig:exp_ef_Gnutella_T}}
	\hspace{-5pt}
	\subfloat[Email-Enron]{\includegraphics[clip,width=0.245\textwidth]{exp/L-effect-E.eps}\label{fig:exp_ef_Email_T}}
	\caption{Effectiveness of varying the random walk length threshold $T$}
	\label{fig:exp_ef_t}
\end{figure}
\fi

%% file: tex/Efficiency.tex
\subsection{Efficiency test}\label{sec_efficiency}
We evaluate the efficiency of different algorithms on Gnutella and Email-Enron datasets. 

\begin{figure*}[t]
	\centering
	\includegraphics[width=0.5\textwidth]{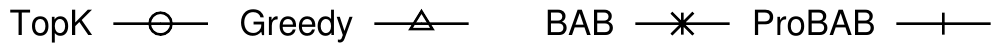}\\
	%\includegraphics[width=1\textwidth]{exp/title5.pdf}
	%\hspace{-10pt}
	\subfloat[Varying $k$]
	{\includegraphics[clip,width=0.3\textwidth]{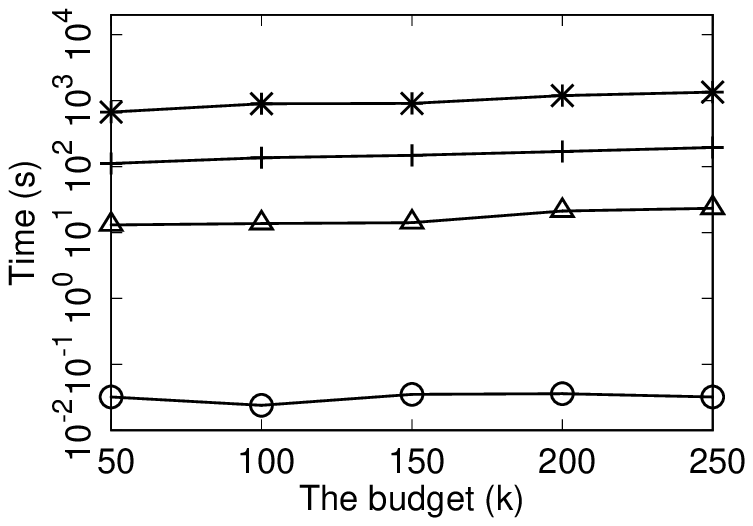}\label{fig:exp_time_Gnutella_B}}
	%\hspace{-10pt}
	\subfloat[Varying $|R|$]
	{\includegraphics[clip,width=0.3\textwidth]{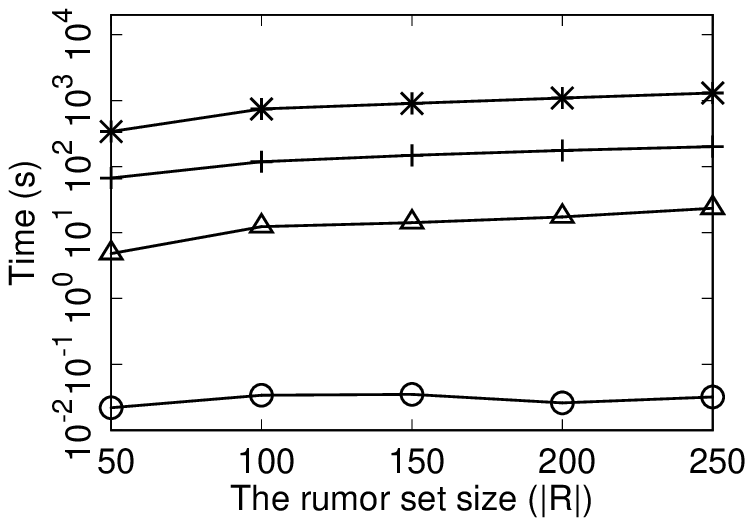}\label{fig:exp_time_Gnutella_R}}
	%\hspace{-10pt}
	\subfloat[Varying $T$]
	{\includegraphics[clip,width=0.3\textwidth]{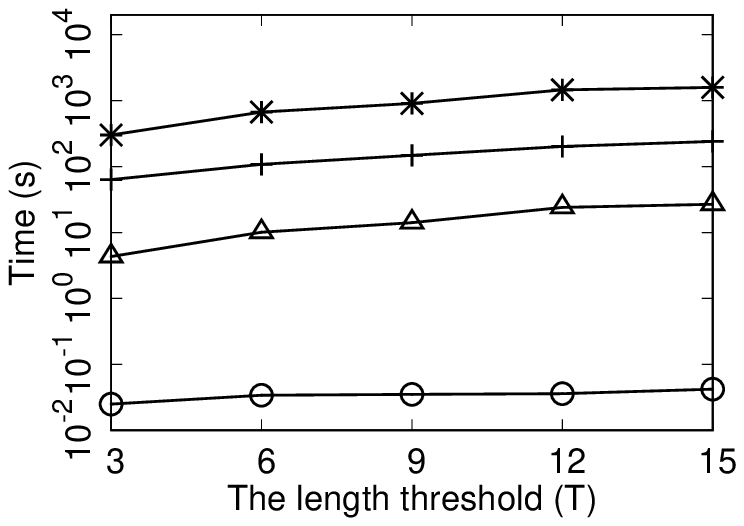}\label{fig:exp_time_Gnutella_T}}
	\caption{Efficiency test on Gnutella}
	\label{fig:para}
\end{figure*}
\begin{figure*}[t]
	\centering
	%\includegraphics[width=0.5\textwidth]{exp/title.pdf}\\
	%\hspace{-10pt}
	\subfloat[Varying $k$]
	{\includegraphics[clip,width=0.3\textwidth]{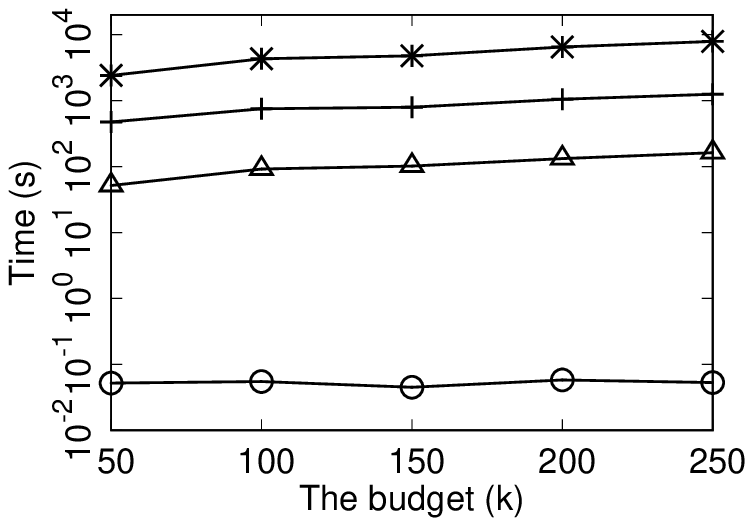}\label{fig:exp_time_Email_B}}
	%\hspace{-10pt}
	\subfloat[Varying $|R|$]
	{\includegraphics[clip,width=0.3\textwidth]{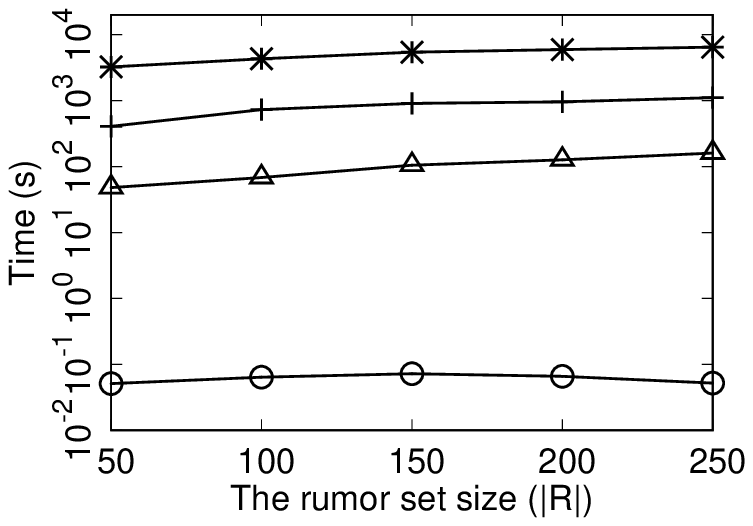}\label{fig:exp_time_Email_R}}
	%\hspace{-10pt}
	\subfloat[Varying $T$]
	{\includegraphics[clip,width=0.3\textwidth]{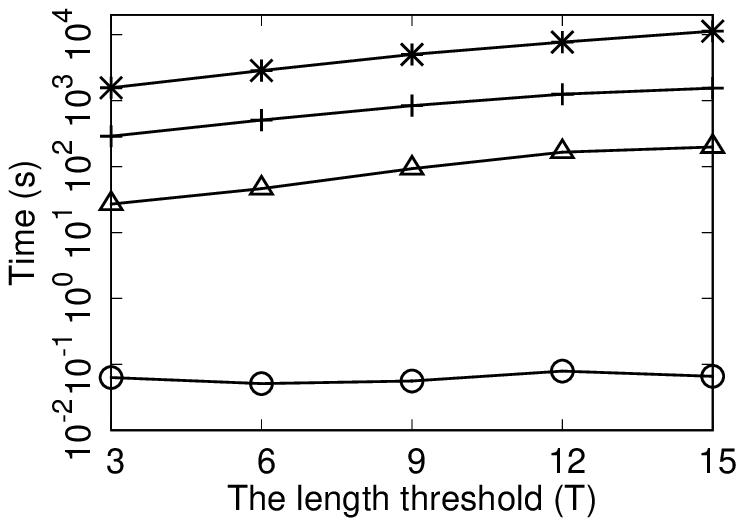}\label{fig:exp_time_Email_T}}
	\caption{Efficiency test on Email-Enron}
	\label{fig:para}
\end{figure*}

\iffalse
\begin{figure}[t]
	\centering
	\includegraphics[clip,width=0.3\textwidth]{exp/title.pdf}\hspace{100pt}
	\subfloat[Gnutella]{\includegraphics[clip,width=0.245\textwidth]{exp/k-time-G.eps}\label{fig:exp_time_Gnutella_B}}
	%\hspace{-5pt}
	\subfloat[Email-Enron]{\includegraphics[clip,width=0.245\textwidth]{exp/k-time-E.eps}\label{fig:exp_time_Email_B}}
	\caption{Efficiency of varying the budget $k$}
	\label{fig:exp_time_k}
\end{figure}
\fi

\noindent
\textbf{Varying the budget $k$.} Figure \ref{fig:exp_time_Gnutella_B}  and Figure \ref{fig:exp_time_Email_B} present the efficiency result when $k$ varies from 50 to 250. We have the following observations. (1) The performance of \samgreedy~and ProBAB is about 2 and 1 orders of magnitude faster than BAB, respectively. (2) The runtime of all methods except \degreeTop~is slowly increasing with the growth of $k$. This is because the increase of $k$ directly causes selecting more nodes to $P$, which leads to an increase in the number of updating the influence block of the remaining node.

\noindent
\textbf{Varying the size of $R$.} Figure \ref{fig:exp_time_Gnutella_R}  and Figure \ref{fig:exp_time_Email_R} show the runtime of all algorithms on Gnutella and Email-Enron, respectively. We can see that the runtime of all methods except \degreeTop~is also slowly increasing when $|R|$ varies from 50 to 250 on all datasets. This is because the influence set $\IS(\ru)$ of $\ru$ is increasing with the growth of $|R|$.

\iffalse
\begin{figure}[t]
	\centering
	%\includegraphics[clip,width=0.5\textwidth]{exp/title.pdf}
	\includegraphics[width=0.5\textwidth]{exp/title.pdf}\\
	\subfloat[Gnutella]{\includegraphics[clip,width=0.245\textwidth]{exp/R-time-G.eps}\label{fig:exp_time_Gnutella_R}}
	\hspace{-5pt}
	\subfloat[Email-Enron]{\includegraphics[clip,width=0.245\textwidth]{exp/R-time-E.eps}\label{fig:exp_time_Email_R}}
	\caption{Efficiency of varying the size of $R$}
	\label{fig:exp_time_r}
\end{figure}
\fi
\noindent
\textbf{Varying the random walk length threshold $T$.} We evaluate the efficiencies of algorithms by varying $T$ from 3 to 15. The result is shown in Figure \ref{fig:exp_time_Gnutella_T} and Figure \ref{fig:exp_time_Email_T}. We can see that all the algorithms except for \degreeTop~scale linearly with respect to $T$, which is because they need to scan more nodes to compute the influence block in each random walk .

\iffalse
\begin{figure}[t]
	\centering
	\includegraphics[clip,width=0.3\textwidth]{exp/title.pdf}\hspace{100pt}
	\subfloat[Gnutella]{\includegraphics[clip,width=0.245\textwidth]{exp/L-time-G.eps}\label{fig:exp_time_Gnutella_T}}
	\hspace{-5pt}
	\subfloat[Email-Enron]{\includegraphics[clip,width=0.245\textwidth]{exp/L-time-E.eps}\label{fig:exp_time_Email_T}}
	\caption{Efficiency of varying the random walk length $T$}
	\label{fig:exp_time_t}
\end{figure}
\fi

%% file: tex/Memory.tex
\subsection{Parameter sensitive test }\label{sec_Parameter} 
\noindent
\textbf{Varying $\beta/\alpha$.} Figure~\ref{fig:exp_pra_t} reports the efficiency and effectiveness of each algorithm when $\beta/\alpha$ is varying. As shown in Figure~\ref{fig:exp_time_Gnutella}, the varying of $\beta/\alpha$ has no impact on the running time of all algorithms. But from Figure~\ref{fig:exp_effect_Gnutella}, we find that the effectiveness of all algorithms is decreasing when the $\beta/\alpha$ varies from $3/7$ to $3/11$. This is because the smaller the $\beta/\alpha$ is, the more times of impression are needed to change a user’s adoption. In particular, with the decrease of $\beta/\alpha$, our solutions outperform \samgreedy~by
70\% to 217\%. Therefore, we choose $\alpha$ = 7 and $\beta$ = 3 as the default setting since our solutions have the smallest advantage of effectiveness for the setting.

%we find: (1) \degreeTop, \samgreedy~and \indexgreedy~have similar memory consumption. It is because they all need to scan the sampling set to calculate the block degree of the nodes. (2) When $T$ is small, the memory consumption of \samgreedyplus~and \indexgreedyplus~is much smaller than that of other methods. This is because only a small number of nodes can reach $R$ when $T$ is small. Therefore, we can reduce the sampling set by eliminating more nodes in pre-sampling stage. But with the growth of $T$, most nodes will reach $R$ in pre-sampling stage, which causes the increasing of the memory consumption of \samgreedyplus~and \indexgreedyplus.

\begin{figure*}[t]
	%\hspace{0.8in}
        %\includegraphics[width=0.35\textwidth]{exp/title.pdf}
	\begin{minipage}[b]{0.48\textwidth} %{0.45\textheight}
		\centering
	\subfloat[Time]{\includegraphics[clip,width=0.5\textwidth]{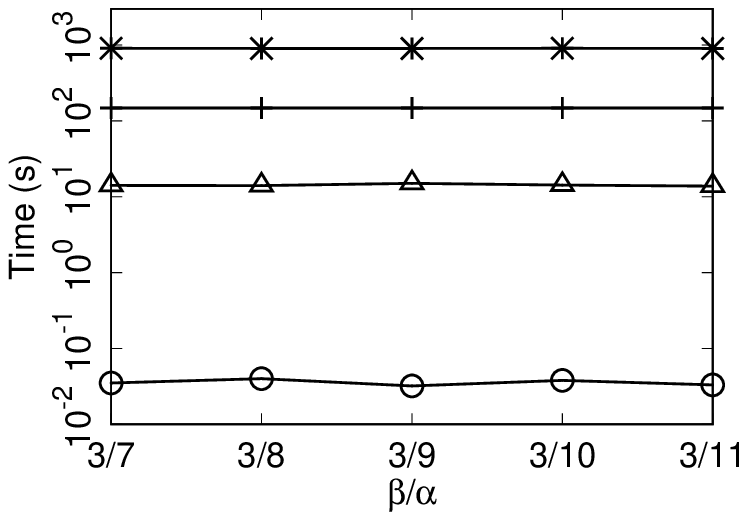}\label{fig:exp_time_Gnutella}}
	\hspace{-5pt}
	\subfloat[Effect]{\includegraphics[clip,width=0.5\textwidth]{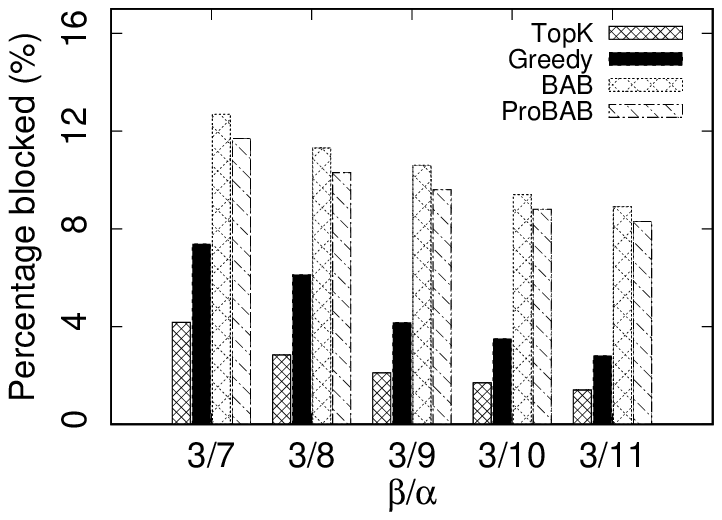}\label{fig:exp_effect_Gnutella}}
	\caption{Varying $\beta/\alpha$ in Gnutella}
	\label{fig:exp_pra_t}
	\end{minipage}
	%\hspace{1in}
	\begin{minipage}[b]{0.48\textwidth}
		\centering
	\subfloat[Time]{\includegraphics[clip,width=0.5\textwidth]{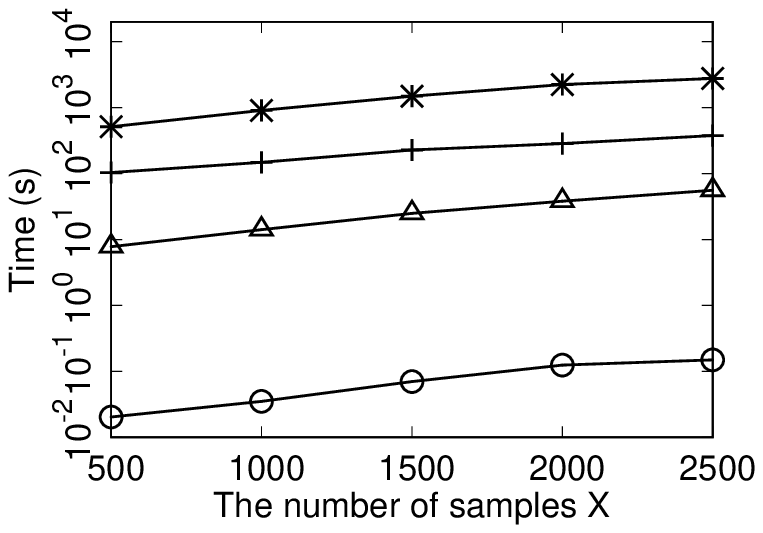}\label{fig:x_time_Gnutella}}
	\hspace{-5pt}
	\subfloat[Effect]{\includegraphics[clip,width=0.5\textwidth]{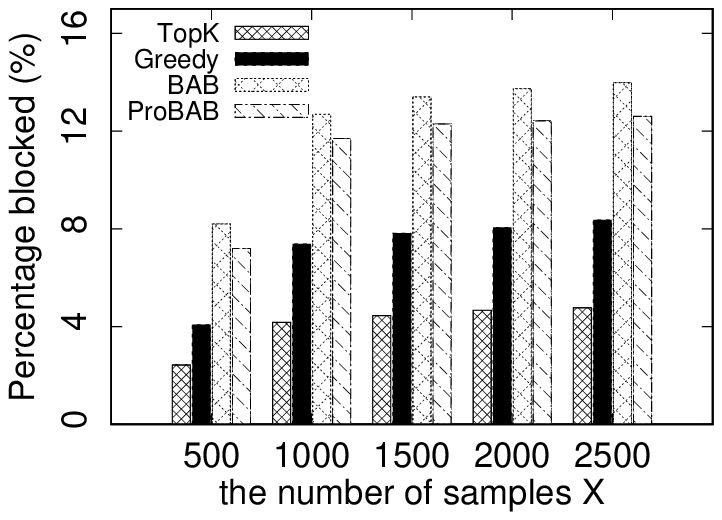}\label{fig:x_effect_Gnutella}}
	%\caption{Varying the number of samples $X$ in Gnutella}
        \caption{Varying $X$ in Gnutella}
	\label{fig:exp_pra_x}
	\end{minipage}
\end{figure*}

% \begin{figure}[t]
% 	\centering
% 	%\includegraphics[clip,width=0.5\textwidth]{exp/title.pdf}
% 	\includegraphics[width=0.35\textwidth]{exp/title.pdf}\\
% 	\subfloat[Time]{\includegraphics[clip,width=0.245\textwidth]{exp/p-time-G.eps}\label{fig:exp_time_Gnutella}}
% 	\hspace{-5pt}
% 	\subfloat[Effect]{\includegraphics[clip,width=0.245\textwidth]{exp/p-effect-G.eps}\label{fig:exp_effect_Gnutella}}
% 	\caption{Varying $\beta/\alpha$ in Gnutella}
% 	\label{fig:exp_pra_t}
% \end{figure}

\noindent
\textbf{Varying the number of samples $X$.} The efficiency and effectiveness of each algorithm when the number of samples $X$ is varying is shown in Figure~\ref{fig:exp_pra_x}. In Figure~\ref{fig:x_time_Gnutella}, the running time of all algorithms increases almost linearly w.r.t. $X$, because all algorithms need to traverse all sampling random walks to calculate the marginal gains or the block degree. From Figure~\ref{fig:x_effect_Gnutella}, we can see that the effectiveness of all algorithms is increasing when the $X$ varies from $500$ to $2500$. But We find that when $X\ge1000$, the changing of effectiveness tends to be stable. Therefore, we choose $X$ = 1000 as the default setting because it reaches an
ideal balance of efficiency and effectiveness.

% \begin{figure}[t]
% 	\centering
% 	%\includegraphics[clip,width=0.5\textwidth]{exp/title.pdf}
% 	\includegraphics[width=0.35\textwidth]{exp/title.pdf}\\
% 	\subfloat[Time]{\includegraphics[clip,width=0.245\textwidth]{exp/x-time-G.eps}\label{fig:x_time_Gnutella}}
% 	\hspace{-5pt}
% 	\subfloat[Effect]{\includegraphics[clip,width=0.245\textwidth]{exp/x-effect-G.eps}\label{fig:x_effect_Gnutella}}
% 	\caption{Varying the number of samples $X$ in Gnutella}
% 	\label{fig:exp_pra_x}
% \end{figure}

\iffalse
\begin{figure*}[t]
	\centering
	%\includegraphics[clip,width=0.5\textwidth]{exp/title.pdf}
	\hspace{-10pt}
	\subfloat[Time]{\includegraphics[clip,width=0.345\textwidth]{exp/p-time-G.eps}\label{fig:exp_time_Gnutella}}
	\hspace{-10pt}
	\subfloat[Effect]{\includegraphics[clip,width=0.345\textwidth]{exp/p-effect-G.eps}\label{fig:exp_effect_Gnutella}}
	\hspace{-10pt}
	\subfloat[Effect]{\includegraphics[clip,width=0.345\textwidth]{exp/p-effect-G.eps}\label{fig:exp_effect_Gnutella}}
	\caption{Varying $\rho$ in Gnutella}
	\label{fig:exp_pra_t}
\end{figure*}
\fi

\begin{figure*}[t]
	%\hspace{0.8in}
        %\includegraphics[width=0.35\textwidth]{exp/title.pdf}
	\begin{minipage}[b]{0.48\textwidth} %{0.45\textheight}
		\centering
	\subfloat[Time]{\includegraphics[clip,width=0.5\textwidth]{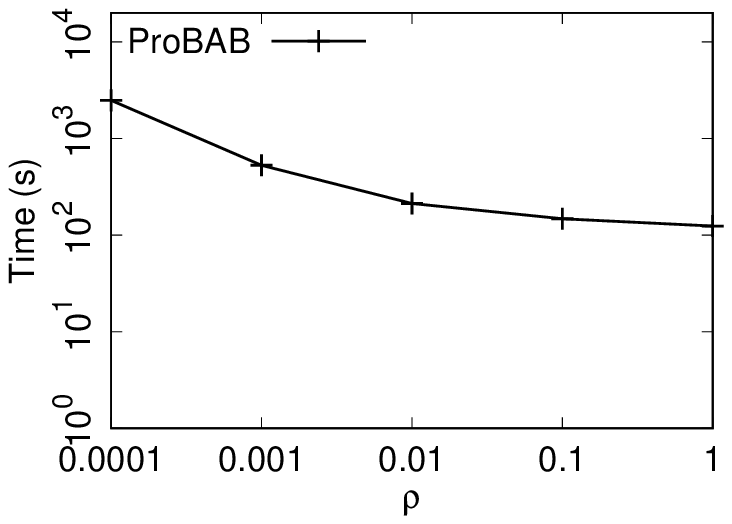}\label{fig:rho_time_Gnutella}}
	\hspace{-5pt}
	\subfloat[Effect]{\includegraphics[clip,width=0.5\textwidth]{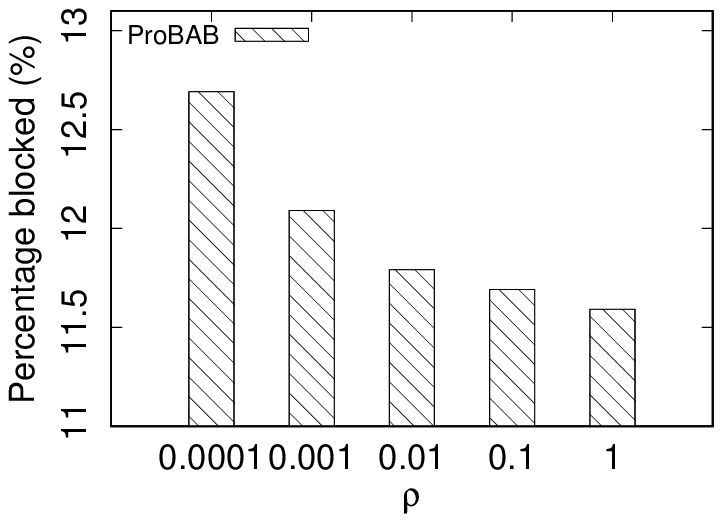}\label{fig:rho_effect_Gnutella}}
	\caption{Varying $\rho$ in Gnutella}
	\label{fig:exp_pra_rho}
	\end{minipage}
	%\hspace{1in}
	\begin{minipage}[b]{0.48\textwidth}
		\centering
	\subfloat[Time]{\includegraphics[clip,width=0.5\textwidth]{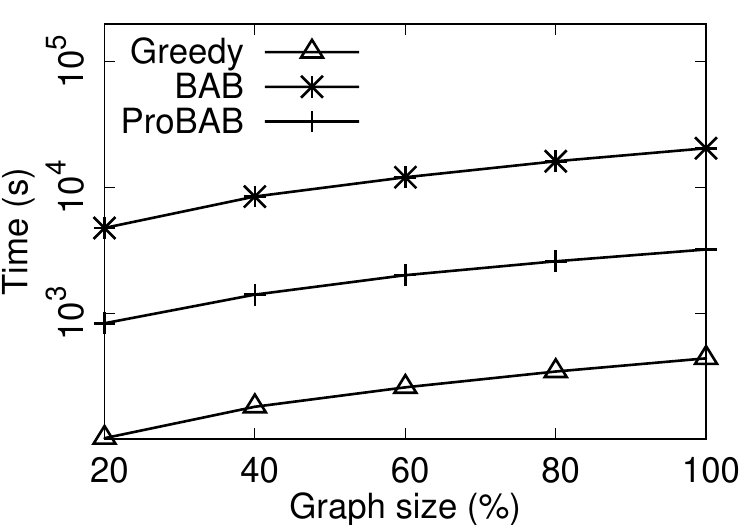}\label{fig:exp_scalability_time}}
	\hspace{-5pt}
	\subfloat[Memory]{\includegraphics[clip,width=0.5\textwidth]{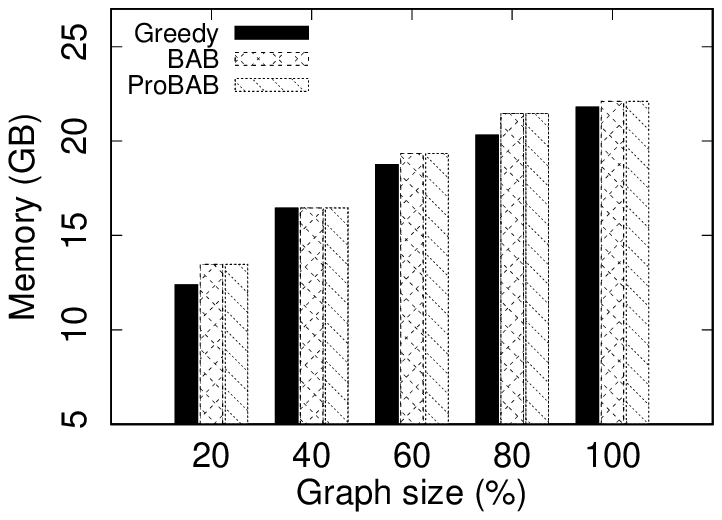}\label{fig:exp_scalability_space}}
	\caption{Scalability test on Gowalla}
	\label{fig:exp_scalability}
	\end{minipage}
\end{figure*}

\noindent
\textbf{Varying $\rho$.} $\rho$ is used to adjust the step distance of decreasing threshold $h$ in Algorithm~4. Figure~\ref{fig:exp_pra_rho} shows the experimental results of varying $\rho$. When $\rho$ increases from 0.0001 to 1, our solutions decrease by at most 10\% in effectiveness but speed up by 1 order of magnitude.
We find that when $\rho\ge$ 0.1, the changing of effectiveness and efficiency tends to be stable. Therefore, we choose $\rho =$ 0.1 as the default setting.

% \begin{figure}[t]
% 	\centering
% 	%\includegraphics[clip,width=0.5\textwidth]{exp/title.pdf}
% 	\subfloat[Time]{\includegraphics[clip,width=0.245\textwidth]{exp/rho-time-G.eps}\label{fig:rho_time_Gnutella}}
% 	\hspace{-5pt}
% 	\subfloat[Effect]{\includegraphics[clip,width=0.245\textwidth]{exp/rho-effect-G.eps}\label{fig:rho_effect_Gnutella}}
% 	\caption{Varying $\rho$ in Gnutella}
% 	\label{fig:exp_pra_rho}
% \end{figure}

%% file: tex/Scalability.tex
\subsection{Scalability test}\label{sec_scalability}
\begin{figure}[h]
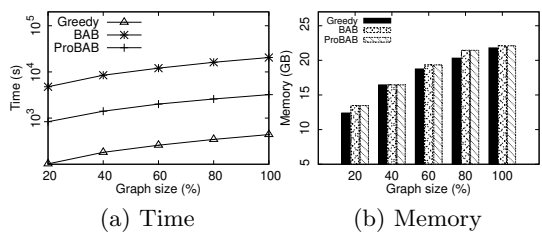

	\centering%\vspace{-0.8cm}
	\subfloat[Time]{\includegraphics[clip,width=0.3\textwidth]{exp/scal-time.eps}\label{fig:exp_scalability_time}}
	\hspace{-5pt}
	\subfloat[Memory]{\includegraphics[clip,width=0.3\textwidth]{exp/scal-mem.eps}\label{fig:exp_scalability_space}}
	\caption{Scalability test on Gowalla dataset}
	\label{fig:exp_scalability}
\end{figure}

This experiment is to evaluate the scalability of \samgreedy~and BAB~when we increase the network size. To vary the network size, we partition Gowalla dataset into five subgraphs, and each of them covers 20\% nodes of the dataset. To avoid smashing the network into pieces, each subgraph is generated by a breadth-first traversal process. %It is worth noting that if one method has more than 25G of memory consumption or runs more than 3000s, we will omit it. 
Figure \ref{fig:exp_scalability} shows the result, and we have the following observations. (1) The performance of \samgreedy~and ProBAB is about 2 and 1 orders of magnitude faster than BAB, respectively. (2) When the graph size is increasing, the memory consumption of \samgreedy, BAB and ProBAB is increasing slowly but no more than 25GB.

%The run time of \samgreedy~and \indexgreedy~increases faster than that of \samgreedyplus~and \indexgreedyplus, with the growth of graph size. This is because the runtime of \samgreedy~and \indexgreedy increases with respect to $n^2$, while the runtime of \samgreedyplus~and \indexgreedyplus increases with respect to $n\log n$. (2) When the graph size is increasing, the memory consumption of \degreeTop, \samgreedy~and \indexgreedy~also increases faster than that of \samgreedyplus~and \indexgreedyplus. As we analyzed in Section~\ref{sec_4:pre-sampling}, \presample~can eliminate some node, which can't reach rumor set $\ru$. Therefore, \presample~can slow down the growth of the memory consumption when the graph size is increasing.

% \begin{wrapfigure}{r}{.5\textwidth}
% 	%\centering
% 	\subfloat[Time]{\includegraphics[clip,width=0.245\textwidth]{exp/scal-time.eps}\label{fig:exp_scalability_time}}
% 	\hspace{-5pt}
% 	\subfloat[Memory]{\includegraphics[clip,width=0.245\textwidth]{exp/scal-mem.eps}\label{fig:exp_scalability_space}}
% 	\caption{Scalability test on Gowalla dataset}
% 	\label{fig:exp_scalability}
% \end{wrapfigure}

%% file: tex/con.tex
\vspace{-0.3cm}\section{Conclusion}\vspace{-0.2cm}
In this paper, we studied the \prob~problem based on a non-submodular influence block model and proved that it is NP-hard to approximate. %More importantly, a simple sampling-based greedy method cannot work well since the logistic-based influence block model is not submodular. 
%Then, we proposed a sampling-based method to compute the upper
%bound by using a dynamic tangent line to tightly bound the actual influence block. By utilizing this upper bound computation method, we built a branch-and-bound framework to solve \prob~problem. 
Then, we proposed a branch-and-bound framework with a sampling-based upper-bound estimation method to solve \prob~problem.
To further improve the efficiency, we optimized our framework with a progressive sampling-based greedy method for upper bound estimation. Lastly, we conducted experiments on real-world datasets to verify the efficiency, effectiveness, and scalability of our methods.\\

%% file: samplepaper.bbl
\begin{thebibliography}{10}
\providecommand{\url}[1]{\texttt{#1}}
\providecommand{\urlprefix}{URL }
\providecommand{\doi}[1]{https://doi.org/#1}

\bibitem{albert2000error}
Albert, R., Jeong, H., Barab{\'a}si, A.L.: Error and attack tolerance of
  complex networks. nature  \textbf{406}(6794), ~378 (2000)

\bibitem{DBLP:conf/wine/BharathiKS07}
Bharathi, S., Kempe, D., Salek, M.: Competitive influence maximization in
  social networks. In: WINE. pp. 306--311 (2007)

\bibitem{DBLP:conf/wine/BorodinFO10}
Borodin, A., Filmus, Y., Oren, J.: Threshold models for competitive influence
  in social networks. In: WINE. pp. 539--550 (2010)

\bibitem{DBLP:conf/www/BudakAA11}
Budak, C., Agrawal, D., {El Abbadi}, A.: Limiting the spread of misinformation
  in social networks. In: WWW. pp. 665--674 (2011)

\bibitem{DBLP:conf/ACMicec/CarnesNWZ07}
Carnes, T., Nagarajan, C., Wild, S.M., van Zuylen, A.: Maximizing influence in
  a competitive social network: a follower's perspective. In: ACMicec. pp.
  351--360 (2007)

\bibitem{feder1985adoption}
Feder, G., Just, R.E., Zilberman, D.: Adoption of agricultural innovations in
  developing countries: A survey. EDCC  \textbf{33}(2),  255--298 (1985)

\bibitem{DBLP:conf/kdd/HabibaYBS08}
Habiba, Yu, Y., Berger{-}Wolf, T.Y., Saia, J.: Finding spread blockers in
  dynamic networks. In: SNAKDD. pp. 55--76 (2008)

\bibitem{hoeffding1994probability}
Hoeffding, W.: Probability inequalities for sums of bounded random variables.
  In: The Collected Works of Wassily Hoeffding, pp. 409--426. Springer (1994)

\bibitem{lancaster1990econometric}
Lancaster, T.: The econometric analysis of transition data. No.~17, Cambridge
  university press (1990)

\bibitem{DBLP:conf/icde/LiYHC14}
Li, R., Yu, J.X., Huang, X., Cheng, H.: Random-walk domination in large graphs.
  In: ICDE. pp. 736--747. {IEEE} Computer Society (2014)

\bibitem{DBLP:journals/pvldb/MoBZP20a}
Mo, S., Bao, Z., Zhang, P., Peng, Z.: Towards an efficient weighted random walk
  domination. PVLDB  \textbf{14}(4),  560--572 (2020)

\bibitem{DBLP:conf/nctcs/MoT0P19}
Mo, S., Tian, S., Wang, L., Peng, Z.: Minimizing the spread of rumor within
  budget constraint in online network. In: NCTCS. vol.~1069, pp. 131--149.
  Springer (2019)

\bibitem{nemhauser1978analysis}
Nemhauser, G.L., Wolsey, L.A., Fisher, M.L.: An analysis of approximations for
  maximizing submodular set functions—i. MP  \textbf{14}(1),  265--294 (1978)

\bibitem{newman2002email}
Newman, M.E., Forrest, S., Balthrop, J.: Email networks and the spread of
  computer viruses. Physical Review E  \textbf{66}(3),  035101 (2002)

\bibitem{palda1965measurement}
Palda, K.S.: The measurement of cumulative advertising effects. The Journal of
  Business  \textbf{38}(2),  162--179 (1965)

\bibitem{spitzer2013principles}
Spitzer, F.: Principles of random walk, vol.~34. Springer Science \& Business
  Media (2013)

\bibitem{taylor2009once}
Taylor, J., Kennedy, R., Sharp, B.: Is once really enough? making
  generalizations about advertising's convex sales response function. Journal
  of Advertising Research  \textbf{49}(2), ~198 (2009)

\bibitem{DBLP:conf/cikm/TripathyBM10}
Tripathy, R.M., Bagchi, A., Mehta, S.: A study of rumor control strategies on
  social networks. In: CIKM. pp. 1817--1820 (2010)

\bibitem{DBLP:journals/www/ZhangBNZMGP19}
Zhang, P., Bao, Z., Niu, Y., Zhang, Y., Mo, S., Geng, F., Peng, Z.: Proactive
  rumor control in online networks. WWW  \textbf{22}(4),  1799--1818 (2019)

\bibitem{DBLP:conf/kdd/ZhangLBMZ19}
Zhang, Y., Li, Y., Bao, Z., Mo, S., Zhang, P.: Optimizing impression counts for
  outdoor advertising. In: SIGKDD. pp. 1205--1215. {ACM} (2019)

\end{thebibliography}
